\documentclass[12pt,letterpaper]{article}
\usepackage{amsmath,amssymb,amsthm,graphicx,a4wide,enumerate,url,tikz}
\usepackage[small,bf]{caption}
\usepackage{subcaption}
\usepackage{colortbl}
\usepackage{color}
\usepackage{physics}
\usepackage{bm}
\usepackage{tikz}
\usetikzlibrary{matrix}
\usepackage{algorithm,algpseudocode}
\usepackage[title]{appendix}
\usepackage{booktabs}
\usepackage{nomencl}

\usepackage{array}
\newcolumntype{C}[1]{>{\centering\arraybackslash}p{#1}}
\usepackage{multirow}

\usepackage[markup=underlined]{changes}

\makenomenclature
\usepackage{etoolbox}
\renewcommand\nomgroup[1]{%
  \item[\bfseries
  \ifstrequal{#1}{O}{Other notations}{%
  \ifstrequal{#1}{P}{Principled Constants}{%
  \ifstrequal{#1}{E}{Empirical Constants}{%
  \ifstrequal{#1}{S}{Principled maps}{%
  \ifstrequal{#1}{F}{Empirical maps}{}}}}}%
]}

\definecolor{darkgreen}{rgb}{0,0.5,0}
\definecolor{purple}{rgb}{1,0,1}

\theoremstyle{plain}

\theoremstyle{definition}

\theoremstyle{remark}

\newcount\Preprint
\newcommand{\Sref}[1]{\ifnum\Preprint=1 \S\ref{#1}\else Section~\emph{\nameref{#1}}\fi}

\usepackage{color,soul}

\definecolor{berkeleyblue}{rgb}{0,0.20,0.38}

\title{Reducing Detailed Vehicle Energy Dynamics to Physics-Like Models}
\author{Nour Khoudari, Sulaiman Almatrudi, Rabie Ramadan, Joy Carpio,\\Mengsha Yao, Kenneth Butts, Alexandre M.~Bayen, Jonathan W.~Lee,\\Benjamin Seibold}
\date{\today}
\begin{document}
\maketitle
\begin{abstract}
The energy demand of vehicles, particularly in unsteady drive cycles, is affected by complex dynamics internal to the engine and other powertrain components. Yet, in many applications, particularly macroscopic traffic flow modeling and optimization, structurally simple approximations to the complex vehicle dynamics are needed that nevertheless reproduce the correct effective energy behavior. This work presents a systematic model reduction pipeline that starts from complex vehicle models based on the Autonomie software and derives a hierarchy of simplified models that are fast to evaluate, easy to disseminate in open-source frameworks, and compatible with optimization frameworks. The pipeline, based on a virtual chassis dynamometer and subsequent approximation strategies, is reproducible and is applied to six different vehicle classes to produce concrete explicit energy models that represent an average vehicle in each class and leverage the accuracy and validation work of the Autonomie software.
\end{abstract}

\section{Introduction}
The accurate quantification of the energy demand of vehicles is of critical importance for many applications, including the design of modern energy-centric intelligent transportation systems. For the quantification of energy demand, given a drive cycle that the vehicle is executing, numerous high fidelity energy modeling and software tools exist, such as Autonomie~\cite{Autonomie_description}, CarSim and TruckSim~\cite{CARSIM}, PSAT~\cite{PSAT}, and VESim~\cite{VESim}. These detailed tools are highly customizable to build specific vehicle models, and they have been validated against real vehicle measurements. So they can, under proper circumstances, serve as suitable proxies for the true real-world energy consumption of vehicles. At the same time, there are several reasons why simpler energy models may be warranted: (i)~Vehicle customization in complex tools requires a significant level of knowledge in vehicle dynamics. (ii)~Many studies in which the collective energy demand of multi-vehicle traffic flow is needed cannot afford the the detailed models' complexity and computational cost, despite parallel computing frameworks that many of these tools provide, such as Autonomie \cite{Autonomie_description, ANL-ML}. (iii)~The models in many of the tools are not publicly available or open source, which may hinder open-source reproducible research. (iv)~The vehicle energy models tend to be complex, thus rendering energy optimization challenging.

For those reasons, in many situations where the energy of a whole traffic stream is to be quantified, significantly simpler models tend to be employed. Many such models (see \Sref{sec:prior_work}) are based on simple physical arguments to describe the energy consumption rate in terms of speed and acceleration, using only basic vehicle characteristics such as frontal area, mass, or tire properties to model aerodynamic drag and rolling friction. However, such simple models are frequently designed completely independently of the detailed models and aforementioned energy tools. Hence, the sophistication, fidelity, validation work, and ongoing developments and updates are not leveraged to the benefit of the much simpler energy models used in numerous multi-vehicle studies that focus on the energy of traffic flow.

In addition, many research efforts focus on devising vehicle controls that optimize energy, for instance \cite{Malikopoulos2012, Lee2012, delle2019feedback, Beaver2019, IshtiaqueMahbub2019}. In situations of a single vehicle, complex energy models can be used. However, recent thrusts focused on making the whole traffic flow of all vehicles on a roadway more energy-efficient by means of properly controlling a few immersed vehicles. For instance, automated vehicles can stabilize uniform flow by dampening stop-and-go waves. This concept of sparse Lagrangian traffic control was first demonstrated experimentally by Stern et al.~\cite{stern2017dissipation}. It has also been captured in simulation is many other works, such as \cite{Cui2017, Milanes2014, Xiao-Yun2015, Mahmassani2001}. As an example, the CIRCLES project \cite{lee2021integrated, Hayat2022holistic, CIRCLES_AMR2022, circles_web} develops such sparse Lagrangian controllers that are (near) energy-optimal, by exploiting reinforcement learning optimization techniques \cite{lee2021integrated} to minimize (under suitable constraints) the energy demand of \emph{all} vehicles on the road. In such settings, energy model frameworks are needed that can capture the effective heterogeneity of vehicle types on real roadways, as well as the fact that the vehicle mix in the bulk flow may not be fully known. Moreover, it is critical that optimization-friendly energy models are employed that average out non-convexity behavior (for instance due to gear switching) of vehicle-specific detailed models, to prevent the optimizer from becoming trapped in local minima that are not practically meaningful. The specifics of such ``convexity'' requirements of energy models are discussed in \Sref{subsec:physic-like-properties}.

This paper addresses these needs by developing a systematic model reduction pipeline that first approximates high-fidelity vehicle models into structurally simpler semi-principled models, and then subsequently approximates those models via simplified fitted functions that can be written via simple mathematical expressions and possess desirable structural properties \cite{circles-energy-models}. For the high-fidelity vehicle models the software Autonomie \cite{AUTONOMIE} is used, based on a representative portfolio of widely used vehicle types (from Compact Sedan to Class8 Tractor, see \Sref{subsec:vehicle_portfolio}). The vehicles considered here are internal combustion engine (ICE) vehicles. However, many aspects of the general model reduction framework are in principle extendable to hybrid and electric vehicles, albeit with important structural adaptations needed such as regenerative braking or battery performance modeling.

\subsection{Prior Work}
\label{sec:prior_work}
Fuel consumption models can be classified in many different ways. For instance, the in-depth literature review \cite{Faris2011} classifies them into modeling categories based on the following: (i)~the scale of the input variables: microscopic, mesoscopic, and macroscopic, (ii)~the formulation approach: analytical, empirical, statistical, and graphical, (iii)~the type of explanatory variables: average speed, instantaneous speed, and specific power models, (iv)~the state variable values: crank-angle resolution-based models and mean value-based models, and (v)~the number of dimensions: zero/one-dimensional/single zone, quasi dimensional, and multi-dimensional/multi-zone. In \cite{Faris2011} the pros and cons of models are discussed comprehensively assuming that the dynamics of a single given vehicle are to be described, without any time constraints. The focus of this work lies on models that yield energy estimates, possibly in real-time applications, for representative average vehicles (within a specific vehicle class), based on instantaneous speed, acceleration, and road grade. To that end, we categorize single-vehicle fuel models into principled models and empirical models, as well an intermediate subcategory of hybrid models to which our own simplification pipeline work belongs: producing a physics-like model that possesses/reproduces key structural aspects of a vehicle's internal dynamics, by fitting data generated by principled models.

\subsubsection{Principled Models}
Principled fuel consumption models use mathematical equations based on physical reasoning that describe the vehicle operation and interactions within its subsystems. For instance, the thermodynamics-based model \cite{Lavoie-Blumberg1980} yields fuel consumption as a function of engine parameters and operating conditions for spark ignition (SI) engines, and had been validated. A detailed but computationally demanding non-linear model of the dynamics of reciprocating engines, involving torsional crankshaft flexibility and its impact on the engine moment of inertia, was proposed in~\cite{METALLIDIS2003723} and used to study the effects of engine misfire.
The physics-based model AMFC~\cite{Khayyam2008} predicts fuel consumption based on the different loads acting on the vehicle including road grade, road-friction, wind-drag, accessories, and mechanical losses. A Bernoulli model of internal combustion engines for vehicle infrastructure \cite{Ni-Henclewood2008} relates engine power and the fuel consumption rate (fuel rate). It was proved reliable against empirical data, however, its complexity rendered it nontrivial to integrate into applications. Under that same category fall also many graphical models like the Simulink-based integrated model of vehicle systems \cite{Assanis2000} which demonstrates the interactions between turbocharged, intercooled diesel engine, driveline, and vehicle dynamics modules. High fidelity is achieved by capturing in particular the thermodynamics of the in-cylinder processes with transient capabilities \cite{Faris2011}. Similar is Autonomie~\cite{AUTONOMIE}, a high-fidelity software developed by Argonne National Laboratory, that estimates fuel consumption of different vehicle classes via Simulink-based components that replicate the internal vehicle subsystems and their input-output interconnection. Autonomie includes an energy library for several powertrain types which provide estimates of energy consumption and other vehicle performance characteristic such as emissions, regenerative braking, losses from braking, aerodynamic drag, or road grade~\cite{AUTONOMIE}. This paper uses Autonomie as the ground truth to build (see \Sref{semiprincipled_pipeline}) and validate (see \Sref{sec:validation}) the developed new models. Other high-fidelity software based on physics-based reasoning include the Vehicle Engine Simulation Model (VESim) \cite{VESim}, which employs a feed-forward simulation with dynamic equations of vehicle sub-system modules, and the Powertrain System Analysis Toolkit (PSAT) \cite{PSAT}, which is a forward-looking model using transient behavior and control systems to generate realistic fuel consumption estimates. Common to most principled models and their implementation into software tools is that they are complex and computationally intensive \cite{lindhjem2004analysis}.

\subsubsection{Empirical Models}
Empirical fuel consumption models are data-driven, based on regression or statistical methodologies applied to processed experimental data, rather than physical reasoning. For example, the Comprehensive Modal Emission Model (CMEM)~\cite{An1997} is based on data collected from 300 real world vehicles tested under a chassis dynamometer. The statistical VT-Micro model~\cite{Ahn-Trani1999, Rakha2000, Ahn2002}, based on chassis dynamometer measurements by Oak Ridge National Laboratory, predicts instantaneous fuel consumption based on instantaneous speed and acceleration, by defining different driving modes (acceleration, idling, cruising, deceleration). VT-Micro does not consider road grade as a direction input, instead it has to be modeled via the acceleration input. The model was extended to very high speed ranges~\cite{Park2010DerivationOR}, as well as to the microscopic emission and fuel consumption (MEF) model~\cite{Lei2010} which also considers memory effects by taking into account accelerations in the recent past. Empirical models are also incorporated in high fidelity simulators that consider the whole vehicle (or its engine) as a black box~\cite{ZHOU2016203}. The VT-Micro model was adopted in the micro-simulation package INTEGRATION, which has been used in many applications~\cite{Yue2008} and whose performance has been demonstrated~\cite{Tapani2005}. Also widely used is the Motor Vehicle Emission Simulator (MOVES)~\cite{lindhjem2004analysis}, a vehicle-based fuel consumption and emission model/simulator developed by the US EPA (United States Environmental Protection Agency). It performs well at estimating fuel rates and emissions, but it is not inexpensive to run~\cite{Faris2011}. Other similar tools/simulators are TruckSim and CarSim \cite{CARSIM}, which use mathematical models for vehicle components and its environment to compute fuel consumption using fitted functions. A fundamental drawback of many empirical models (and software that incorporate them) is that any changes/updates to the physical characteristics of a vehicle or its engine would require the very costly experimental re-collection and re-fitting of data.

\subsubsection{Hybrid Models}
There are many models that lie in between the two categories above, by employing the complexity and physics nature of principled models to characterize the internal system of vehicles, and at the same time draw benefit from the data-driven nature of empirical models. An example of such a hybrid model is the Vehicle Transient Emissions Simulation Software (VeTESS)~\cite{Pelkmans2004}, which was built using measurements from real-world traffic and chassis dynamometer. The model inputs are the vehicle speed and road grade, and outputs are emission and fuel rates. Through reverse simulation, VeTESS determines engine speed and torque. It is based on the assumption that under transient conditions, engine speed and output torque are dictated by a series of quasi-steady states of the dynamic engine behaviour. The Virginia Tech Comprehensive Power-Based Fuel Consumption Model (VT-CPFM)~\cite{edwardes2014virginia, edwardes2015modeling} considers the effects of aerodynamic drag, rolling resistance, inertial effects, and road grade. The model \cite{edwardes2014virginia}, for diesel and hybrid buses, was calibrated via the Orange County bus cycles dynamometer test. Similarly, \cite{park2013virginia} used VT-CPFM for diesel buses, using EPA real-world city and highway fuel economy ratings for calibration. The models developed herein are also examples of hybrid models. Leveraging the principled nature of Autonomie, maps are extracted and fitted to polynomial models, see \Sref{semiprincipled_pipeline}.

\subsection{Approach and Novelty}
The energy modeling procedure proposed in this paper uses a principled high-fidelity energy modeling tool as a baseline (Autonomie is used here, but the methodology could also be extended to other tools). The baseline software is then used to generate data in the form of a \emph{Virtual Chassis Dynamometer} (VCD). Combined with a few other vehicle parameters that are extracted directly from Autonomie (such as mass, road load, upshift gear schedule), fitted mappings between fuel rate and engine speed and engine torque are constructed. These are incorporated into simpler and more direct physical models to yield stateless physical models that we refer to as the \emph{semi-principled models}. These models are then further simplified in such a way that explicit formulas arise (not requiring any data tables) that are fast to evaluate and allow direct and convenient energy optimization. The resulting models therefore leverage much of the flexibility and accuracy of Autonomie, but are conceptually and structurally much simpler and much less computationally demanding. Another key contribution of this work is that all constructed models are carefully validated on a variety of EPA and UNECE (United Nations Economic Commission for Europe) drive cycles, and their accuracy is demonstrated.

\subsection{Structure and Content}
In the next sections of this paper, a detailed description of the energy modeling pipeline is presented; starting from the derivation of the MATLAB-based semi-principled model (\Sref{semiprincipled_pipeline}) from Autonomie using the VCD (\Sref{subsec:VCD}) to a simpler physics-like fitted Simplified model (\Sref{sec:simplified_pipeline}) that is easier to computationally execute. The modeling pipeline also gives emphasis on the special treatments applied for heavy-duty (\Sref{subsec:special_treatment_HDV}) and manual transmission vehicles (\Sref{subsec:manual_trans}), explains certain choices in our modeling/fitting to achieve physics-like models, and describes the region of feasible speed/acceleration/grade combination acceptable for the models (\Sref{subsec:feasible_region}). In the latter sections, there is a description of the representative vehicles used to model an ``average'' vehicle in each of the six vehicle classes that were used as a demonstration in our portfolio. Results of the fitting parameters for all six Simplified models is shown followed by validation results of our models versus Autonomie measurements as ground truth under standard drive cycles on flat roads (\Sref{subsec:validation_flatroads}) and under our construction of constant nonzero road grade drive cycles (\Sref{subsec:validation-with-grade}). The study is concluded with a thorough discussion of the validation results and their connection to some modeling choices that ensured a balance between simplicity and accuracy.

\section{From Detailed Vehicle Energy Models to Semi-Prin{\-}cipled Models}
\label{semiprincipled_pipeline}
The first step in the model reduction pipeline is to build semi-principled energy models from vehicle-specific detailed energy models. The reduction process described here works in principle for any software that estimates vehicle energy consumption and runs VCD experiments provided that it has access to the necessary parameters/maps to be extracted and used.

In this study, Autonomie Rev 16SP7 was used to evaluate energy parameters of conventional powertrains for the desired vehicle class. All the baseline vehicle model configurations used as reference and assumptions in the Autonomie compiled vehicles are provided in \cite{DOE_Report}. Autonomie has more than 100 powertrain configurations for conventional, hybrid (series, parallel, and split), and battery electric types with low- and high-level controls available for most of the powertrains. There are also several mathematical models for the hardware components of the system (called ``plant models'') with more than 100 initialization modes that can be used to customize a specific vehicle model. A vehicle configuration model is a MATLAB- and Simulink-based model that represents the input-output behavior of a vehicle as influenced by the driver model, environment model, high-level vehicle powertrain controller, and vehicle propulsion architecture. 
In the vehicle propulsion architecture section are the plant and controller models for the engine, clutch, gearbox, chassis, and some mechanical and electrical accessories of the vehicles. Pre-defined light-, medium-, and heavy-duty class vehicle models that are comparable to vehicles currently on the market are available in the Autonomie library. Thus, Autonomie vehicle model simulation results are considered representative of a vehicle class's energy consumption as validated on relevant drive cycles.

In this section, we describe the structure of the semi-principled model, followed by a detailed description of the process for extracting the necessary parameters and maps via VCD to be used in the semi-principled model.

\subsection{Structure of the Semi-Principled Model}
\label{structure_semi_model}
The semi-principled model takes as an input the instantaneous vehicle speed $v$, acceleration $a$, and road grade $\theta$, and it outputs, as direct functions $f = f(v,a,\theta)$, the instantaneous engine speed $N$, engine torque $T$, fuel rate $f$, transmission output speed $N_{\text{output}}$, wheel force $F_{\text{wheel}}$, wheel power $P_{\text{wheel}}$, engine power $P_{\text{engine}}=NT$, gear, and feasibility of the given $(v,a,\theta)$ triplet.
This is a fundamental structural simplification over Autonomie, where the output signals (such as the fuel rate $f(t)$) depend on the input signals ($v(t)$ and $\theta(t)$) non-locally in time.

It should be noted that for a given angle of road inclination $\gamma$ (measured in $\text{rad}$), the corresponding road grade is $\theta = \tan\gamma$. Moreover, the gravitational acceleration along the road, as it is relevant for energy models, scales with $\sin\gamma$. However, for small inclination angles ($\gamma \le 0.06\text{rad}$, as it is the case on US highways), the difference between the quantities $\gamma\approx\sin{\gamma} \approx \tan{\gamma}$ is negligible. Hence, in this work we use these quantities interchangeably.

We denote the constructed model ``semi-principled'', because it has a physics-based part based on known vehicle-specific quantities (such as the vehicle mass $m_{\text{vehicle}}$ and general vehicle mass $m^k_{\text{general}}$ depending on gear, road load coefficients $R_{\text{a}}$, $R_{\text{r}}$, and $R_{\text{g}}$, final drive ratio $d_{\text{r}}$, tire radius $r_{\text{tire}}$, and gear ratio $g_{\text{r}}$, all described in \Sref{subsubsec:parameters_autonomie}), but it also relies on fitted maps that are obtained from the VCD (such as vehicle speed to engine speed, vehicle speed and wheel force to engine torque, and engine speed and torque to fuel rate, described in \Sref{subsubsec:empirical_maps}). The following description of the structure of the semi-principled model uses notations of parameters and maps extracted from the VCD. A detailed definition and description of how such quantities are tuned or fitted is provided in \Sref{subsec:extract_prameters_and_maps_from_autonomie}. Any quantity that depends on the gear ($k^\text{th}$ gear) carries a superscript $k$.

For all the calculated quantities listed below, refer to the pseudo-code in Algorithm~\ref{alg:semi-prinipled-model} for the detailed formulas used in the implementation of the semi-principled model. Given vehicle speed $v$, acceleration $a$, and road grade $\theta$, we calculate, for each gear $k$: (i)~the transmission output $N_{\text{output}}$, (ii)~the commanded wheel force $F^k_{\text{wheel}} = m^k_{\text{general}} a + R_{\text{a}}  v^2 + R_{\text{r}} v + R_{\text{g}} +   m_{\text{vehicle}}\sin(\theta)g$, where $g=9.81\text{m}/{s^2}$ is the gravitational constant, (iii)~the corresponding wheel power $P^k_{\text{wheel}}$, and (iv)~the wheel torque $T^k_{\text{wheel}}$. Note that we use the generalized vehicle mass $m^k_{\text{general}}$ (defined in \Sref{subsubsec:parameters_autonomie}) in the calculation of the wheel force. This ensures that the force required to accelerate the vehicle does depend on the gear choice because the internal components' rotation speeds depend on the gear.

Next, we determine the pedal angle $\alpha^k$, as it is defined by Autonomie, as the wheel torque over the maximum wheel torque $T_{\text{wmax}}(v)$ at the commanded speed $v$. This pedal angle is then used to determine the gear. Gear scheduling involves a critical simplification step: in line with real vehicles, Autonomie has a hysteresis between upshifting and downshifting, and the selected gear depends on the recent history of driving states. In contrast, the semi-principled model is designed to be local in time, that is, the gear choice must be determined based solely on the instantaneous speed, acceleration, and road grade. We conduct this simplification step by using only the upshift map $K_{\text{upshift}}(\alpha,v)$ to define feasible gears, and then later select the gear based on energy-optimality. The results in \Sref{sec:validation} demonstrate that this approximation (particularly simplifying the downshifts events) tends to not result in significant errors in fuel rates. Below, Autonomie's gear choice will be denoted as $g^k_{\text{Autonomie}}$. Next we determine upper limits for the wheel force and wheel torque. The maximum wheel torque $T^k_{\text{wmax}}$ (and consequently the maximum wheel force $F^k_{\text{wmax}}$) is calculated using $T_{\text{wmax}}(v,k)$, the maximum wheel torque map per gear at the given speed. 

The simplest modeling of vehicle dynamics is to assume that the torque converter bypass clutch is always locked. However, this simplification results in underpredictions of the fuel rate in lower gears because in real vehicles an open torque converter bypass facilitates vehicle launch and mitigates driveline vibration at the expense of increased fuel consumption. Thus, in our model the torque converter is assumed to have an open bypass clutch in first gear only following the current industry modeling trend. With the assumption of having an open torque converter in first gear, the engine speed $N^k$ and engine torque $T^k$ are calculated using the fitted mappings from transmission output speed and wheel force to engine operating condition $N(N_\text{output},F_\text{wheel},k)$ and $T(N_\text{output},F_\text{wheel},k)$, respectively. Those maps are defined by the fitted coefficients from the earlier discussed fitting routines of transmission output speed and wheel force to engine speed/torque data. For the first gear ($k=1$), we treat the open torque state separately by taking as inputs to the fitted maps $N_{\text{output}}:=\min(N_{\text{output}},N_{\text{max}}/g^1_{\text{r}})$ and
$F^1_{\text{wheel}}:=\min(F^1_{\text{wheel}},F^1_{\text{max}})$ as the transmission output speed value and the wheel force, respectively. Note that this modeling choice assumes steady state of the torque converter which will lead to an underestimation of engine torque in the first gear torque converter impeller at high and medium acceleration rates when the torque converter is in a transient state, which we correct for by adding a term representing an average underestimation value calculated using $T_{\text{correction}}(a)$ the open engine torque correction values map (defined in \Sref{subsubsec:parameters_tuning}) evaluated at the commanded acceleration. For all gears, we limit the engine torque and engine speed from below by $T_{\text{min}}$ and $N_{\text{min}}$.

Finally, the fuel consumption $f^k$ at each gear $k$ is calculated using the fitted fuel rate map $f_{\text{r}}(N,T)$ at the calculated $N^k$ and $T^k$. The calculated fuel rate is capped from below by zero to avoid negative values. If the fuel-cut condition is satisfied, that is, $v>v_{\text{c}}$ and $F^k_{\text{wheel}}<F_{\text{wc}}$, then the fuel rate is set to zero as well.  

We discourage very low engine speeds in cases where the vehicle is not coming to a stop by assigning a penalty value $g^k_{\text{penalty}}$ for mismatching Autonomie's gear choice. Penalties are defined as some measure of badness and violations to viable engine operating conditions that will manifest in the gear choice output of our model: if $k>1$ and $g^k_{\text{Autonomie}}<k$ for a non-fuel-cut case, we assign a penalty for that particular gear $k$, namely $g^k_{\text{penalty}}=k-g^k_{\text{Autonomie}}$. To ensure that the acceleration request can be delivered by the powertrain at the current vehicle speed and road grade we assign penalty values. At high wheel force requests: if $F^k_{\text{wheel}}>F^k_{\text{max}}$, we assign a penalty for that particular gear $k$, namely $F^k_{\text{penalty}}=F^k_{\text{wheel}}-F^k_{\text{max}}$. We can find the maximum engine torque $T^k_\text{max}$ using the maximum engine torque map $T_{\text{max}}(N)$ at the calculated engine speed $N^k$. To ensure that the engine is actually able to produce the requested torque and the engine speed is below the maximum speed we assign a penalty value for exceeding the maximum engine torque $T^k_\text{max}$, namely $T^k_{\text{penalty}} = T^k-T^k_\text{max}$, and for exceeding the maximum engine speed $N_\text{max}$, namely $N^k_{\text{penalty}} = N^k-N^k_\text{max}$, in a non-fuel-cut case for that particular gear $k$.

With the engine conditions, fuel consumption, and penalties calculated at each gear, we choose the gear with the lowest sum $f^k+w_TT^k_\text{penalty}+w_NN^k_\text{penalty}++w_FF^k_\text{penalty}+w_gg^k_{\text{penalty}}$ where the weights are chosen as follows: $w_T = 10\text{g}/(\text{s}\cdot\text{Nm})$, $w_N = 10\text{g}/\text{rad}$, $w_F = 100\text{g}/(\text{s}\cdot\text{N})$, and $w_g = 100\text{g}/\text{s}$. To determine whether a commanded combination of speed, acceleration, and road grade is feasible for the engine to handle, we define a feasibility marker with a value equal to the penalty value $E^k_{\text{penalty}}$ at that chosen gear. If that marker value is nonzero, then the commanded speed $v$, acceleration $a$, and road grade $\theta$ are not feasible. If we are in state of braking, that is, $F_{\text{wheel}}<0$, the gear is chosen based on the gear downshifting map $K_{\text{downshift}}(v)$, the construction of which is described in \Sref{subsubsec:parameters_tuning}. This ensures that shifting during deceleration generates a coherent shift map. Finally, for the tuned gear choice and the commanded speed $v$, acceleration $a$, and road grade $\theta$, we return the corresponding outputs of the semi-principled model. If the vehicle is operating at zero speed and acceleration, the vehicle is assumed to be idling with the engine operating at the tuned parameters $N_{\text{min}}$, $T_{\text{min}}$ and consuming $f_{\text{idle}}$. The remaining outputs are then calculated from these three engine idle operating conditions (described in \Sref{subsubsec:parameters_tuning}). The reasoning for this treatment is to improve the model accuracy because the idle vehicle condition (zero speed and acceleration) is a common evaluation point.

\begin{algorithm*}[!ht]
\caption{Semi-principled model pseudo-code}
\label{alg:semi-prinipled-model}
\begin{algorithmic}
\State Given $v$, $a$, and $\theta$
\For{all gears $k$}
    \State $N_{\text{output}}=\frac{d_{\text{r}} v}{r_{\text{tire}}}$\Comment{Transmission Output}
    \State $F^k_{\text{wheel}} = m^k_{\text{general}} a + R_{\text{a}}  v^2 + R_{\text{r}} v + R_{\text{g}} +   m_{\text{vehicle}}\sin(\theta)g$;\Comment{Wheel Force}
    \State $P^k_{\text{wheel}} = F^k_{\text{wheel}} v$;\Comment{Wheel Power}
    \State $T^k_{\text{wheel}} = F^k_{\text{wheel}}r_{\text{tire}}$\Comment{Wheel Torque}
    \State $\alpha^k = \frac{T^k_{\text{wheel}}}{T_{\text{wmax}}(v)}$\Comment{Pedal Angle}
    \State $g^k_{\text{Autonomie}}=K_{\text{upshift}}(\alpha^k,v)$ \Comment{Autonomie gear choice}
    \State $T^k_{\text{wmax}}=T_{\text{wmax}}(v,k)$
    and $ F^k_{\text{wmax}} = T^k_{\text{wmax}}/ r_{\text{tire}}$ \Comment{Maximum Wheel Toque and Force}
    \State $N^k = \max(N_{\text{min}},N(N_\text{output},F_\text{wheel},k))$ \Comment{Engine Speed}
    \State $T^k = \max(T_{\text{min}},T(N_\text{output},F_\text{wheel},k))$ \Comment{Engine Torque}
    \State $T^k_{\text{max}}=T_{\text{max}}(N^k)$ \Comment{Maximum Engine Torque}
    \If{$k=1$} 
\State $N_{\text{output}}=\min(N_{\text{output}},\frac{N_{\text{max}}}{g^1_{\text{r}}})$ and $F^k_{\text{wheel}}=\min(F^1_{\text{wheel}},F^1_{\text{max}})$
\State $T^1=T^1 + T_{\text{correction}}(\max(a,0))$ \Comment{Torque Correction}
    \EndIf
    \State $f^k=\max(0,f_{\text{r}}(N^k,T^k))$ \Comment{Fuel Rate}
    \If{$v>v_{\text{c}}$ and $F^k_{\text{wheel}} <F_{\text{wc}}$}
        $f^k=0$ \Comment{Fuel-cut condition}
    \ElsIf{$k>1$ and $g^k_{\text{Autonomie}}<k$ }
        assign penalty $g^k_{\text{penalty}}$  \Comment{Gear penalty}
    \ElsIf{$N^k>N_{\text{max}}$} 
        assign penalty $N^k_{\text{penalty}}$ \Comment{Max Engine Speed penalty}
    \ElsIf{$T^k>T^k_{\text{max}}$}
       assign penalty to $T^k_{\text{penalty}}$ \Comment{Max Engine Torque penalty}
    \ElsIf{$F^k_{\text{wheel}}>F^k_{\text{max}}$}
        assign penalty to $F^k_{\text{penalty}}$  \Comment{Max Wheel Force penalty}
    \EndIf
\EndFor
\State Choose $k$ with least weighted sum $f^k+w_TT^k_\text{penalty}+w_NN^k_\text{penalty}+w_FF^k_\text{penalty}+w_gg^k_{\text{penalty}}$
\end{algorithmic}
\end{algorithm*}

\subsection{Generation of Data via the VCD}
\label{subsec:VCD}
In this section, the generation of the semi-principled model is described step-by-step, beginning with obtaining baseline data for the unmodified template Autonomie vehicle model on select standard drive cycles to the description of required modifications on the vehicle model to run the VCD.  
In this context, a VCD is a gear-by-gear test pattern programmed in MATLAB to run the Simulink model of an appropriately modified vehicle from Autonomie by inputting a driving pattern and extracting vehicle measurements such as fuel rate and engine operating conditions that will be used to build the empirical maps described in \Sref{subsubsec:empirical_maps}. The exact modifications on the template Autonomie vehicle model may slightly change for different types of vehicles depending on their transmission type or driver model, for example, in order to force the vehicle to execute the test patterns gear-by-gear. However, the general procedure to generate the semi-principled model using a VCD is summarized below.

\medskip\noindent\textbf{Step 1: Generate the baseline Autonomie data.}
The unmodified template Autonomie vehicle model is run on the selected standard drive cycles, described in \Sref{subsec:drive_cycles}. The resulting data from these runs are then used as a baseline against which the semi-principled (and other) models are validated.

\medskip\noindent\textbf{Step 2: Modify the template Autonomie model.}
The template vehicle structure in Autonomie is cloned and customized to make it compatible with running the VCD. Target plants and controllers are modified in Simulink, while Autonomie initialization files are directly edited. The following steps are conducted to prepare the vehicle model for the VCD:
\begin{itemize}
    \item \textbf{Environment model:} The VCD test pattern is defined in the initialization file of the environment model for each gear $k$. The test speed-profile is defined as piece-wise constant from the gear's minimum road speed $v^k_{\text{min}}$, to the gear's maximum road speed $v^k_{\text{max}}$, in increments of $0.1 \text{m}/\text{s}$. $v^k_{\text{min}}$ and $v^k_{\text{max}}$ are defined as follows:
    \begin{align*}      v^k_{\text{min}} &= \max\left\{\frac{N_{\text{min}} r_{\text{tire}}}{g^k_{\text{r}} d_{\text{r}}}, \ 0\text{m}/\text{s}\right\}\;,\\
        \quad
        v^k_{\text{max}} &= \min\left\{\frac{N_{\text{max}} r_{\text{tire}}}{g^k_{\text{r}} d_{\text{r}}}, \ 34\text{m}/\text{s}\right\}\;,
    \end{align*}
    
    where $N_{\text{min}}$, $N_{\text{max}}$, $r_{\text{tire}}$, $g^k_{\text{r}}$ and $d_{\text{r}}$ are principled constants extracted from Autonomie and described in \Sref{subsubsec:parameters_autonomie}. At each constant speed, the applied test pedal angle is defined as piece-wise constant $[0\% ,  100\%]$ in increments of $2\%$, and each test speed and pedal angle is run for $10 \text{s}$ to ensure steady state operation of the vehicle model. The desired torque converter bypass clutch status for the VCD is also defined in the initialization file of the environment model for each gear, as described in \Sref{structure_semi_model}. In the environment plant model, a closed-loop feedback controller is added which modifies a virtual road grade applied to counteract the prescribed pedal angle applied in the VCD, defined by the test pattern, so that the vehicle speed matches the test target speed.
    
    \item \textbf{Driver model:} The driver's acceleration and braking commands in the driver plant model are replaced by forcing the accelerator pedal and target vehicle speed to execute the test pattern that is run on the VCD.

    \item \textbf{Wheel plant:} The additional load induced by the virtual road grade is bypassed to avoid additional driving force in the force calculation at the wheel. This is achieved by replacing the value of the road grade cosine block in the ``Force Calculation'' block to 1 always.
    \item \textbf{Gearbox controller:} The gear transient and demand controllers are set to ``steady gear'' and ``no gear shifting'' conditions while still allowing the normal engine mode determination. Gear choice and torque converter bypass clutch status are set to follow the gear-by-gear test pattern defined in the environment model initialization file.
    \item \textbf{Clutch/torque converter:} Restrict the torque converter operational mode to only be in transient, steady state or locked, such that idle mode is not allowed. 
\end{itemize}

\medskip\noindent\textbf{Step 3: Simulate the modified model}
The resulting modified model is then run on any generic drive cycle in Autonomie to produce a MATLAB data file that will include all principled constants and maps described in \Sref{subsubsec:parameters_autonomie} and \Sref{subsubsec:parameters_maps}, 
as well as a Simulink model of the modified vehicle, ready to run the VCD.

\medskip\noindent\textbf{Step 4: Run the VCD}
The just generated Simulink model is now run via a MATLAB script and the operating conditions needed to build the empirical maps, described in \Sref{subsubsec:empirical_maps}, are captured for each (gear, speed, pedal angle) test pattern pairing, $8 \text{s}$ after continuously applying it, to ensure steady vehicle operation.

\subsection{Semi-principled Model Parameters and Maps}
\label{subsec:extract_prameters_and_maps_from_autonomie}
The semi-principled model uses several parameters and maps that describe the vehicle and engine characteristics. The parameters can be classified into two categories: (a)~\textbf{principled constants} that are known from the Autonomie model, and (b)~\textbf{empirical constants} that are not explicitly given but instead are extracted/tuned from the typical/average behavior that the given Autonomie model exhibits when run on standard drive cycles. The maps can be categorized as: (c)~\textbf{principled maps} that are extracted directly from Autonomie, and (d)~\textbf{empirical maps} that are fit to the VCD data.

\subsubsection{Principled Constants}
\label{subsubsec:parameters_autonomie}
The parameters assigned from the Autonomie model are: $m_{\text{vehicle}}$ vehicle mass ($\text{kg}$), $m^k_{\text{general}}$ generalized vehicle mass per gear ($\text{kg}$) (defined by adding a mass value to $m_{\text{vehicle}}$ to account for the drive-line inertia $I^k_{\text{D}}$, such that $m^k_{\text{general}} = m_{\text{vehicle}} + I^k_{\text{D}}/r_{\text{tire}}$) \cite{VehDynamics}, $r_{\text{tire}}$ tire radius ($\text{m}$), road load ($\text{N}$) parameters (three different types: (i)~$R_{\text{a}}$ air resistance depending on the size/shape of the vehicle and air drag, (ii)~$R_{\text{r}}$ rolling resistance depending on nature of road, tires, and weight of the vehicle, and (iii)~$R_{\text{g}}$ frictional load), $d_{\text{r}}$ final drive ratio (number of rotations in transmission output needed to turn the wheels once), $g_{\text{r}}$ gear ratios (ratio of transmission input speed to output speed), 
 $N_{\text{max}}$ maximum engine speed ($\text{rad}/\text{s}$), and $N_{\text{min}}$ engine idle speed or the minimum engine speed ($\text{rad}/\text{s}$).

\subsubsection{Empirical Constants}
\label{subsubsec:parameters_tuning}
Automated procedures are devised to extract the empirical constants. First, the considered Autonomie model is run on all drive cycles that are representative for the vehicle at hand (see \Sref{subsec:drive_cycles}).

The \textbf{minimum engine torque} $T_{\text{min}}$ ($\text{Nm}$) and \textbf{idling fuel rate} $f_{\text{idle}}$ ($\text{g}/\text{s}$) are determined as the median torque and mean fuel rate, respectively, for all data points in the drive data that satisfy the following conditions: (a)~the vehicle speed is below $0.1\text{m}/\text{s}$, while also (b)~the rate of change of torque is below $0.01\text{Nm}/\text{s}$ in magnitude (that is, it is essentially constant in time), for all data points in the interval of $[-1\text{s},+1\text{s}]$.

Fuel-cut is instantiated precisely when two criteria are met: the speed $v$ is above a \textbf{fuel-cut speed} $v_\text{c}$ and the wheel force $F_{\text{wheel}}$ is below a \textbf{fuel-cut wheel force} $F_{wc}$. These two threshold parameters are independently determined as follows. 

The \textbf{fuel-cut speed} $v_{\text{c}}$ ($\text{m}/\text{s}$) is determined as the $1^\text{st}$ percentile for all data points which satisfy: (a)~the fuel rate is below $0.05\text{g}/\text{s}$, and (b)~the vehicle speed is above $1\text{m}/\text{s}$. The $1^\text{st}$  percentile is used instead of the minimum to establish robustness with respect to a few outliers in the data. 

The \textbf{fuel-cut wheel force} $F_{\text{wc}}$ ($\text{N}$) is determined as the $95^\text{th}$  percentile wheel force for the data points satisfying the same criteria used for fuel-cut speed estimation. The Autonomie vehicle model does not apply fuel-cut immediately when the vehicle starts braking but allows the engine to stay engaged for a short period to ensure a state of braking, and remains in fuel-cut until braking is no longer applied. So in addition to allowing for outliers, the choice of the $95^\text{th}$ percentile is critical to mitigate/average Autonomie’s hysteresis.

The downshifting vehicle speeds used to describe the \textbf{downshifting map} $K_{\text{downshift}}(v)$ from speeds to gear. Those vehicle speeds at which downward gear shifting during braking occur are determined as the median speeds for all downshifting events in the data, for each gear-pair, ($k_{\text{before}},k_{\text{after}}$), where $k_{\text{after}} = k_{\text{before}}-1$.  Note that the downshifting vehicle speeds are generally not relevant for the actual fuel consumption as they tend to arise during zero or idle fuel injection rates due to slowing down or braking. Yet, to establish a nice agreement of gear choice patterns between the semi-principled model and the full Autonomie model, these values are extracted. The downshifting vehicle speeds define the cut-off points of a piece-wise constant vehicle speed to gear mapping for downshifting while braking.

The \textbf{open torque converter correction for first gear} $T_{\text{correction}}(a)$ ($\text{Nm}$) is determined as follows. Two regions of low, $a<1 \text{m}/\text{s}^2$,  and high, $a>2 \text{m}/\text{s}^2$, acceleration are found for which (a)~the vehicle is in first gear with open torque converter clutch, and (b)~the torque converter is in steady state mode. For each of the two regions, the acceleration and torque underestimation error is averaged over three samples in the region to give two pairs of $(a,T_{\text{correction}})$. In addition, the origin point $T_{\text{correction}}(0) = 0$ is added to the low and high acceleration estimated torque error to construct the mapping by fitting a linear function $T_{\text{correction}}(a)$ for first gear. The reasoning for this is because when the vehicle is accelerating, some of the power generated by the engine is used to accelerate the torque converter impeller, a behavior which is not captured in the constant speed VCD described above. While this correction resolves the issue of underestimating the engine torque during acceleration in first gear with a steady state open torque converter, it over-corrects the torque for the region before reaching steady state, known as the transient torque converter state. However, the effect of the transient state over-correction is assumed to have a small effect on the fuel consumption estimation due to its short interval, an assumption which the validation results confirm.

\subsubsection{Principled maps}
\label{subsubsec:parameters_maps}
Principled maps are directly extracted from Autonomie results data file, as described in step 3 of \Sref{subsec:VCD}, and stored as arrays (discrete evaluations) to be interpolated when used in the semi-principled model. These maps are: $K_{\text{upshift}}(\alpha,v)$ the automatic gear upshift map for vehicles with an automatic transmission, that is, a map from pedal angle $\alpha$ and speed to gear, $V_{\text{upshift}}(\alpha,k)$ and $V_{\text{downshift}}(\alpha,k)$ the manual upshift and downshift maps, respectively, for manual transmission vehicles ($\text{m}/\text{s}$), that is, maps from pedal angle and gear to upshift and downshift speeds, $T_{\text{max}}(N)$ maximum engine torque map ($\text{Nm}$), that is, a map from engine speed to maximum engine torque, $T_{\text{wmax}}(v)$ maximum wheel torque ($\text{Nm}$), that is, a map from speed to maximum allowed wheel torques, and $T_{\text{wmax}}(v,k)$ maximum wheel torque by gear ($\text{Nm}$), that is, a map from speed and gear to maximum wheel torque. Note that in the semi-principled model, $T_{\text{wmax}}(v,k)$ will be used to calculate the maximum wheel torque (and consequently the maximum wheel force) while $T_{\text{wmax}}(v)$ will be used to calculate the pedal angle to be consistent with Autonomie and to get a good estimation of Autonomie's gear choice result when using the upshift map $K_{\text{upshift}}(\alpha,v)$.

\subsubsection{Empirical maps}
\label{subsubsec:empirical_maps}
The semi-principled model relies on some fitting routines of Autonomie’s VCD data into maps. Those routines are generally chosen to be least square fittings of polynomial classes of functions with the same choice of polynomial degree used across all vehicle classes. In addition, those polynomials are chosen to have the lowest degree because using higher degree polynomials in the fitting routines can add unwanted model complexity, introduce non-typical trends due to over-fitting, and lead to unreasonable values due to extrapolation outside the data range. 

The extracted maps to be fitted are:

\textbf{Engine speed and engine torque to fuel rate:} a bi-variate polynomial function $f_{\text{r}}(N,T)$ ($\text{g}/\text{s}$) of degree 2 in engine speed and degree 3 in engine torque is fitted to the data generated by the VCD.

\textbf{Transmission output speed and wheel force to engine speed for each gear} $N(N_{\text{output}},F_{\text{wheel}},k)$ ($\text{N}$): For the first gear with open torque converter bypass clutch, the data are fitted using a polynomial that is degree 2 in wheel force and degree 3 in transmission output speed. For all higher gears where the torque converter bypass clutch is locked, the engine speed data are observed to be linear and depend only on the rotational speed, so it is fitted using a degree 1 polynomial in the transmission output speed. 

\textbf{Transmission output speed and wheel force to engine torque for each gear} $T(N_{\text{output}},F_{\text{wheel}},k)$ ($\text{Nm}$): For the first gear with open torque converter bypass clutch, the data are fitted using a continuous piece-wise linear fit. We start with a degree 1 polynomial fit in both transmission output speed and wheel force as an initial guess for the piece-wise linear least squares fitting routine to ensure that it converges. For all higher gears where the torque converter bypass clutch is locked, the engine speed data are observed to be linear, so it is fitted using a degree 1 polynomial in both transmission output speed and wheel force.

\subsection{Special Treatment of Manual Transmission}
\label{subsec:manual_trans}
Manual transmission vehicles require different gear scheduling logic to discourage upshifting at low pedal angles. The gear scheduling for manual transmission is achieved by utilizing the principled gear shifting maps extracted from Autonomie. After calculating the pedal angle $\alpha^k$, a shift speed $v^k_{\text{upshift}}$ is found from the manual upshift mapping $V_{\text{upshift}}(\alpha,k)$. A small gear penalty proportional to the difference between the vehicle speed and the upshift speed is assigned if the vehicle speed is below $v^k_{\text{upshift}}$ and the vehicle is not in fuel-cut. This penalty is similar to the one imposed on automatic transmission vehicles when $g^k_{\text{Autonomie}}<k$. After calculating the fuel rate of each gear, the gear penalties, and any engine penalties due to exceeding maximum force, torque or speed, the gear $k_\text{opt}$ with the minimum sum of fuel rate and penalties is chosen. One of the characteristics of both $V_{\text{upshift}}(\alpha,k)$ and $V_{\text{downshift}}(\alpha,k)$ is having a flat shift speed for low pedal angle up to a defined $\alpha_{\text{s}}$. This is utilized to improve gear scheduling; that is, if  $\alpha^{k_\text{opt}}<\alpha_{\text{s}}$, assign the gear $k$ given by reverse looking up the gear from $V_{\text{downshift}}(\alpha,k)$ as the tuned gear choice, otherwise, assign $k_\text{opt}$. The calculated output corresponding to the tuned gear choice is outputted as described for the automatic transmission vehicles and the input $(v,a,\theta)$ is considered feasible if a gear choice results in no engine penalties.

\section{From Semi-Principled Models to Simplified Physics-like Fitted Models}
\label{sec:simplified_pipeline}

\subsection{Motivation for Models Without Gear-Shift Jumps}
\label{subsec:model_motivation}
\begin{figure}
    \centering 
    \includegraphics[width=0.9\linewidth]{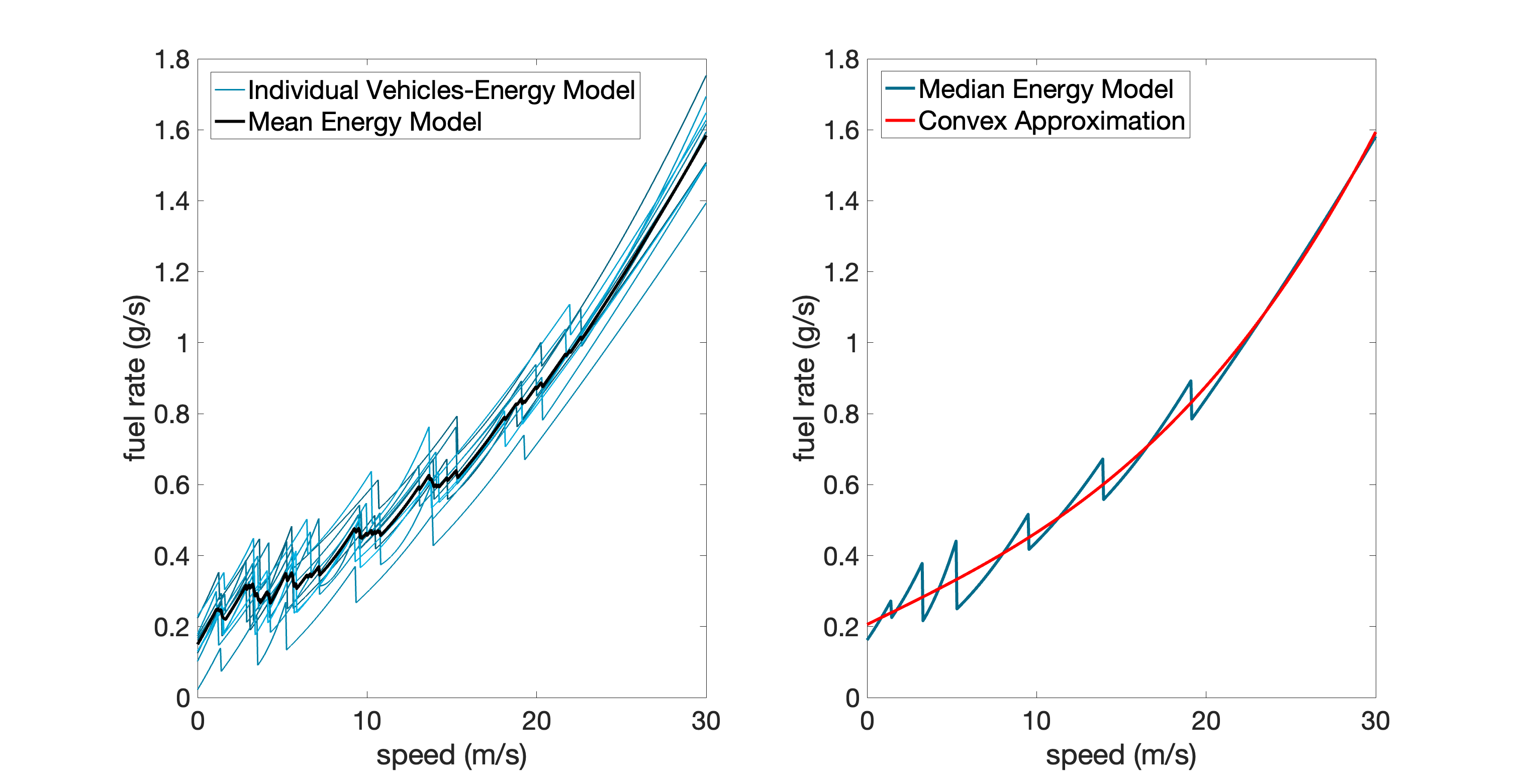}
    \caption{The fuel rate ($\text{g}/\text{s}$) as a function of speed ($\text{m}/\text{s}$) at zero acceleration and road grade of: (left) the semi-principled model of 11 different makes and models of a Midsize SUV (cyan) compared to their average (black), and (right) the semi-principled model of a Midsize SUV vehicle ``in the middle" of vehicles on the market (blue) compared to its corresponding continuous fitted simplified model (red).}
    \label{Mean_Median}
\end{figure}

The second step in the model reduction pipeline is the construction of physics-like models that are locally continuous and averaging out gear shifting behavior that is represented by discontinuities in the semi-principled models. There are multiple reasons for this additional fitting step. First and foremost, there is simplicity: the fitted models will be represented via a few simple mathematical expressions and parameters, and thus are straightforward to communicate and store, and extremely fast to evaluate (a significant advantage when using such models in many-vehicle scenarios).
However, there is also a secondary point in favor of producing fitted models that average out gear-shifting, as follows. In real traffic flow, the collection of vehicles of a given class (like Midsize Sedan) is composed of many different makes and models, and thus the effective energy demand of this collection of vehicles is given by an average (weighted by the penetration rates of different makes and models) of the vehicle-specific energy functions. Because different vehicles have different gear-switching points, the resulting \emph{mean energy model} has many discontinuities, each of which is much smaller than the jumps for each individual vehicle. Figure~\ref{Mean_Median} left depicts such a portfolio of vehicle energy models (light blue curves), and the resulting mean energy model (black curve).

In contrast, what is available via the Autonomie software is a representative specific vehicle of the given class, ``in the middle'' of vehicles on the market \cite{ANL}. We denote the resulting energy model, shown in Figure~\ref{Mean_Median} right (blue curve), as the \emph{median energy model}. The question is now: in applications where the mean energy is desired (such as heterogeneous traffic flow), how can the median model be leveraged to approximate the mean? As one can see in Figure~\ref{Mean_Median} right, a smooth and convex approximation (red curve) serves as a reasonable approximation to the desired mean model. Note that Figure~\ref{Mean_Median} serves only as a motivation, thus displaying only the cruising energy profiles on flat roads. Below we provide a robust fitting procedure on the full phase space.

\subsection{Simplified Model with Desirable Physics-like Properties}
\label{subsec:covex_model}
After having constructed the semi-principled models, we conduct a further fitting step motivated by the idea in \Sref{subsec:model_motivation} and develop a simplified fuel consumption model for each vehicle class. In this approach, we favor the simplicity and interpretability of the models, and therefore we limit the simplified model inputs to vehicle speed, acceleration, and road grade.

The aim is to construct a model that is physics-like and polynomial in spirit. The reason behind such choice is to avoid having optimization problems with solutions exploiting overfitting artifacts. The choice of a function with polynomial terms is motivated by the physical laws that govern the power demand needed to overcome friction forces \cite{Galvin2017, Gong2018}. Therefore, we choose to fit the semi-principled model of each vehicle class to a function $f_{\text{s}}(v,a,\theta)$ of the form
\begin{equation*}
    f_{\text{s}}(v,a,\theta) = \max\left\{\ell(v,a,\theta), \, f_\text{p}(v,a,\theta)\right\}\;,
\end{equation*}
where
\begin{align*}
\ell(v,a,\theta) &=
  \begin{cases}
        \beta & \text{if $v\leq v_\text{c}$}\;, \\
        0 & \text{if $v> v_\text{c}$ and $a<a_c(v,\theta)$}\;,
  \end{cases}
\quad\text{and} \\
f_\text{p}(v,a,\theta) &=  C(v) + P(v)a + Q(v) (a_+)^2 + Z(v)\theta\;.
\end{align*}
This function has 19 parameters (degrees of freedom) that are fitted for each class of vehicles based on data collected from the corresponding semi-principled models. The function $f_\text{p}(v,a,\theta)$ represents fuel rate, and is a polynomial in speed $v$, acceleration $a$, and road grade $\theta$. The function $\ell(v,a,\theta)$ is effectively a lower bound on the amount of fuel consumed by the vehicle: it is zero for high speed due to vehicles enacting a fuel-cut upon deceleration, and it is a constant-free parameter $\beta$ for low velocities. The speed at which a fuel-cut kicks in is another free parameter $v_\text{c}$, and the corresponding acceleration values at which fuel-cut is activated is determined by the polynomial function $a_c(v,\theta)$:
\begin{equation*}
  a_c(v,\theta) = a_0 + a_1v + a_2\theta + a_3v^2 + a_4v\theta\;.
\end{equation*}
For $\theta=0$ and $a=0$, the function $f_\text{p}(v,0,0) = C(v)$ is a cubic polynomial in $v$, shown below. This form is motivated by physics models where $c_0$ ensures that even at zero speeds some fuel will be consumed; $c_1$ and $c_2$ terms contribute to the fuel consumption due to internal and rolling friction forces, respectively; and the $c_3$ term contributes to the fuel consumption due to the air-drag force. When $\theta = 0$, $f_\text{p}(v,a,0)$ is a bivariate polynomial of degree $3$ that caps the effects of the $a^2$ term when $a<a_{\text{min}}$. The reason for that is to avoid the increasing contributions of the $a^2$ terms to the fuel consumption $f_\text{p}(v,a,0)$ as the acceleration decreases below the critical value $a_{\text{min}}(v)$. Note that for simplicity, we could set $a_{\text{min}}=0$, but that would introduce a model with a discontinuous second derivative in $a$ exactly at $a=0$. So instead, we favor to have this discontinuity in the second derivative at exactly the minimum with respect to $a$ while ensuring that $f_\text{p}(v,a,0)>0$ is increasing in $a$ by constraining the minimum $f_\text{p}(v,a_{\text{min}}(v),0)$ to be above zero. These functions are defined as follows:
\begin{align*}
C(v) &= c_0 + c_1 v + c_2 v^2 + c_3 v^3\;,\\
P(v) &=  p_0  + p_1 v + p_2 v^2\;,\\
Q(v) &=  q_0 + q_1 v\;,
\end{align*}
\begin{equation*}
a_+ = \max\{a,a_{\text{min}}(v)\}\;,~\text{and} \quad a_{\text{min}}(v)=-\frac{P(v)}{2Q(v)}\;.
\end{equation*}
For simplicity, we lastly assume that for a fixed speed, the rate at which the fuel consumption increases as a function of $\theta$ is independent of the vehicle acceleration. In that case, fuel consumption changes linearly with $\theta$, and for a given road grade, the rate at which the fuel consumption increases is quadratic as a function of $v$:
\begin{equation*}
Z(v) =  z_0 + z_1 v + z_2 v^2\;.
\end{equation*}
The $z_1$ term is motivated by physics and contributes to fuel consumption due to weight force exerted at the given road grade. The $z_2$ term is not motivated by physics but added to get a better fit of the semi-principled model data, and $z_0$ is added as a lower order term to make the model more general.

The fitting is to be understood in an $L^2$ sense, and all implementations are approximations via quadrature on regular grids. While the fitting could in principle be done in a single step by giving equal weights over a specific grid in $v\times a\times\theta$ phase space, we favor a fitting procedure in multiple steps, giving special attention to scenarios where certain variables vanish because they play singularly special roles: idling ($v=0$), cruising ($a=0$), and flat roads ($\theta=0$). The five-step fitting procedure is:
\begin{itemize}
    \item \underline{Step 1:} Construct the piece-wise constant function $(v,a,\theta)$ by first finding $v_\text{c}$ and $\beta$ from the semi-principled model. $v_\text{c}$ is obtained using the bisection method applied to the semi-principled model as a function of $v$ and at $a=-3\text{m}/\text{s}^2$, $\theta =0 \text{rad}$. The semi-principled data points $f(v,-3,0)$ are collected  over $100$ equispaced grid points of $v \in [0\text{m}/\text{s}, 35\text{m}/\text{s}]$. $\beta$ is found by substituting $v=0 \text{m}/\text{s}$, $a=-3\text{m}/\text{s}^2$, and $\theta =0 \text{rad}$ into the semi-principled model. $a=-3\text{m}/\text{s}^2$ is chosen as a strong enough deceleration level that guarantees the occurrence of the fuel-cut vehicular behavior. $a_0$, $a_1$, $a_2$, $a_3$, and $a_4$ can be found by fitting $a_c(v,\theta)$ to the acceleration values at which the semi-principled fuel rate $f(v,a,\theta)$ hits zero. Those values are obtained using bisection methods applied to the semi-principled model as a function of $a$ and over $50\times50$ equispaced grid points for $v \in [v_\text{c}+1\text{m}/\text{s}, 35\text{m}/\text{s}]$ and $\theta\in[-0.03\text{rad},0\text{rad}]$.
    
    \item \underline{Step 2:} Find the parameters $c_0$, $c_1$, $c_2$, and $c_3$ by fitting the cubic polynomial $f_\text{p}(v,0,0)$ to data generated from the semi-principled model at $a=0\text{m}/\text{s}^2$ and $\theta =0 \text{rad}$ for $v \in [0\text{m}/\text{s}, 35\text{m}/\text{s}]$. The data are generated on $100$ equispaced data grid points, and the domain $[0\text{m}/\text{s}, 35\text{m}/\text{s}]$ is chosen to be consistent with the domain on which the semi-principled model was obtained, while ignoring any data points that are flagged infeasible when generated in the semi-principled model. The fitting is done by minimizing the sum of the squared differences between the cubic polynomial $f_\text{p}(v,0,0)$ and the data, while imposing four constraints; namely that all coefficients must be non-negative to guarantee convexity.
    
    \item \underline{Step 3:} Find the parameters $p_0$, $p_1$, and $p_2$ by fitting the function $C(v)+P(v)a$ to data generated from the semi-principled model at $\theta =0 \text{rad}$ and for $v \in [0\text{m}/\text{s}, 35\text{m}/\text{s}]$ and $a \in [-0.3\text{m}/\text{s}^2, 0.3 \text{m}/\text{s}^2]$. The data are generated on $100\times 100$ equispaced grid points, and the acceleration domain is chosen to capture the slope in $a$ that is more dominant for accelerations around 0 while neglecting the strong quadratic behavior that kicks in for higher speeds. In this fitting step, $c_0$, $c_1$, $c_2$, and $c_3$ are treated as constants (found in Step 2), and we use only feasible and non fuel-cut data points corresponding to $f(v,a,0)\geq\beta$ when $v<v_\text{c}$ or $f(v,a,0)>0$ when $v\geq v_\text{c}$. The fitted parameters are constrained to be no-negative to guarantee convexity.
    
    \item \underline{Step 4:} Find the parameters $q_0$ and $q_1$ by fitting the function $f_\text{p}(v,a,0)$ to data $f(v,a,0)$ generated from the semi-principled model at $\theta =0 \text{rad}$, and for $v \in [0\text{m}/\text{s}, 35\text{m}/\text{s}]$, and $a \in [0\text{m}/\text{s}^2, 3 \text{m}/\text{s}^2]$. The data are generated on $100 \times 100$ equispaced grid points, and the acceleration domain is chosen to capture the quadratic behavior in $a$, using only feasible data points and non fuel-cut data points. In this fitting step, $c_0$, $c_1$, $c_2$, $c_3$, $p_0$, $p_1$, and $p_2$ are treated as constants (found in Steps 2 and 3). The fitting is done by minimizing the sum of squared differences between the polynomial $f_\text{p}(v,a,0)$ and the data $f(v,a,0)$, while imposing two constraints; namely that both parameters must be non-negative.
    
    \item \underline{Step 5:} Find the parameters $z_0$, $z_1$, and $z_2$ by fitting $Z(v)\theta$ to the difference of the data generated from the semi-principled model $f(v,0,\theta)$ and the data generated by the fitted model $f_\text{p}(v,0,0)$ for $v \in [0\text{m}/\text{s}, 35\text{m}/\text{s}]$ and $\theta \in [0\text{rad}, 0.03\text{rad}]$. The data are generated on a $100 \times 100$ equispaced grid points in the $(v,\theta)$ domain, and the road grade domain is chosen so that the relevant drive cycles are still feasible for most vehicles at the maximum road grade in the domain. We ignore data corresponding to negative road grades due to symmetry in most vehicles and to avoid any data outliers that stagnates in the semi-principled model for negative road grades due to minimum engine torque restrictions. While fitting, we capture only feasible data points and non-fuel-cut data points, and we impose three constraints; namely that $z_0$, $z_1$, and $z_2$ must be non-negative.
\end{itemize}
The units commonly associated with the fuel rate function $f_{\text{s}}(v,a,\theta)$ are $\text{g}/\text{s}$, where $\text{g}$ stands for $\text{grams}$. See Table~\ref{table:summary-of-fitted-parameters} for a summary of the units associated with the fitted parameters of this convex model for a portfolio of six vehicle classes.

\subsection{Special Treatment of Medium- and Heavy-Duty Vehicles}
\label{subsec:special_treatment_HDV}
The fitting process described in the above section can be applied to any vehicle class corresponding to the types: light- medium- and heavy-duty vehicles. However, the proposed ranges above on which we carry the fitting should be modified for medium- and heavy-duty vehicles. Such vehicles do not exhibit the option to fuel-cut, therefore for a given vehicle class, if the fuel-cut speed in Step 1 is approximated to be the maximum speed $v_\text{c}=35\text{m}/\text{s}$, this implies that it is a medium- or heavy-duty vehicle. This would introduce additional small valued data points in the fitting process that are usually flagged as fuel-cut data points in light-duty vehicles and are excluded while fitting for negative acceleration values. To avoid including those data points for medium- and heavy-duty vehicles, we restrict the acceleration range in Step 3 to be $a\in[0\text{m}/\text{s}^2,0.3\text{m}/\text{s}^2]$. In addition, the maximum road grade for which most drive cycles corresponding to heavy-duty vehicles remain feasible is generally less than those corresponding to light-duty vehicles. Thus, for medium- and heavy-duty vehicles we restrict the road grade range in Step 5 to be $\theta\in[0\text{rad},0.02\text{rad}]$. Finally, since medium- and heavy-duty vehicles do not exhibit fuel-cut, $\beta$ is not relevant, as the lower bound function $\ell(v,a,\theta)$ is non-constant and varies with $v$. Thus, we fit a linear function $\ell(v,a,\theta)=h_0+h_1v$ to the feasible data generated from the evaluations of the semi-principled model $f(v,-3,0)$ over 100 equispaced grid points of $v\in[0\text{m}/\text{s},35\text{m}/\text{s}]$, for $a = -3 \text{m}/\text{s}^2$ and $\theta = 0 \text{rad}$.

\begin{figure*}
    \centering 
    \includegraphics[width=0.9\linewidth]{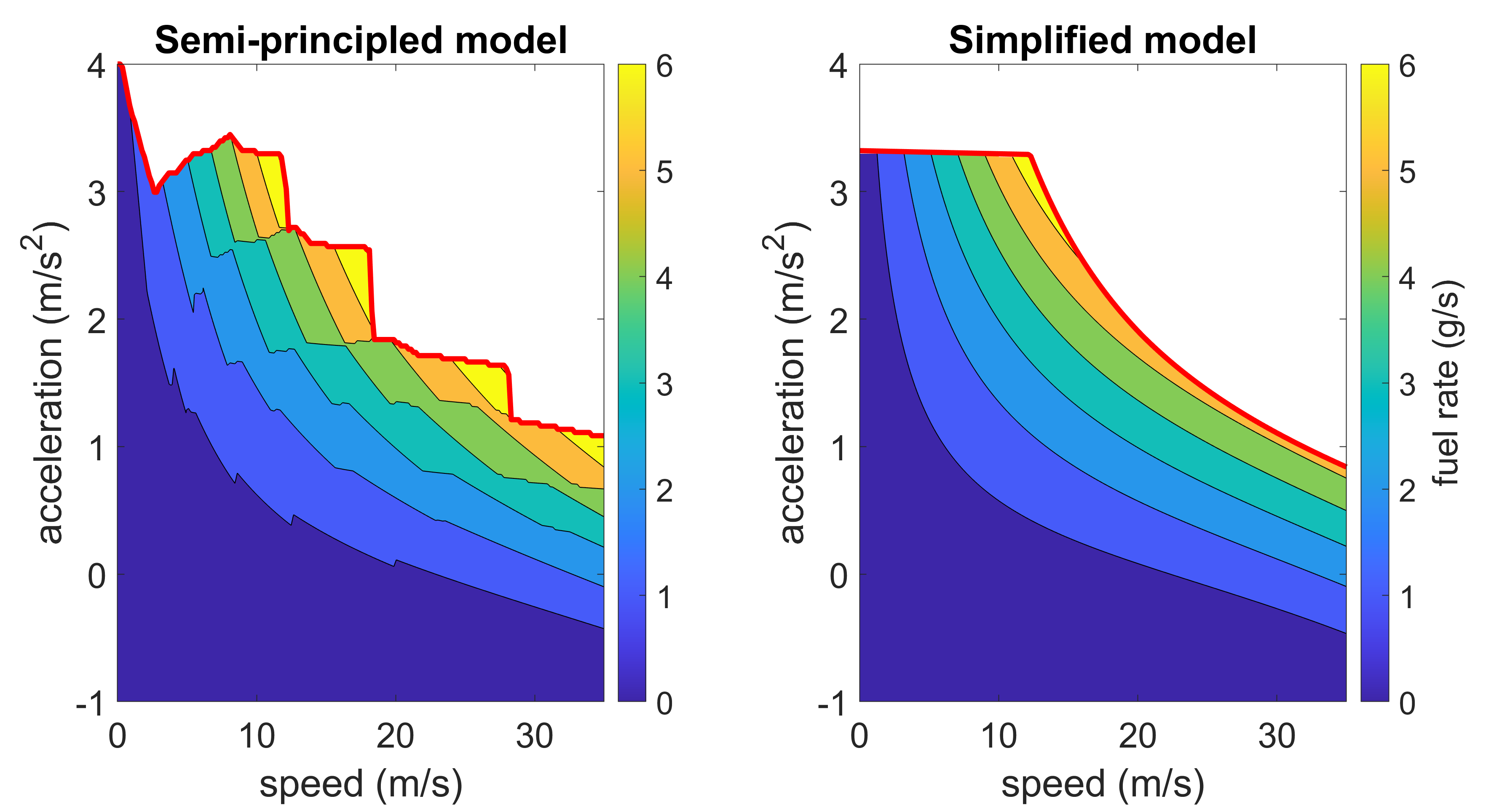}
    \caption{Compact Sedan contours of fuel rates within the feasible region in the ($v$,$a$)-space at 0.01rad road grade for semi-principled model (left) and simplified model (right). The curve in red represents the upper boundary of the feasibility region of each model.}
    \label{fig_feasible_upperbound_SPM_SM}
\end{figure*}

\subsection{Approximation of the Feasible Region}
\label{subsec:feasible_region}
The upper bound of the feasible region of a vehicle in the ($v$,$a$,$\theta$)-space is defined to be the maximum acceleration $a_{max}(v,\theta)$ at which a vehicle's engine could still function properly for a given combination of speed $v$ and road grade $\theta$. This upper bound has discontinuities in the semi-principled model; however, since our final goal is to build a smooth simplified model, we choose to approximate this upper boundary in the simplified model with piecewise continuous functions. This choice is rooted in the physics of combustion engines, and the function that is used to fit the upper boundary of the semi-principled model is of the form
\begin{equation*}
a_{\text{max}}(v,\theta) = \min\left(b_1, \frac{b_2}{v}-b_3v^2\right)
-\min\left(b_4, b_5+b_6v\right)\theta\;.
\end{equation*}
This function has six parameters that are fitted for each vehicle class. For $\theta=0$, the function $a_{\text{max}}(v,0)$ can be understood from true physics by assuming that the power at the wheel is constant for all relatively large speeds. This is because as the number of discrete gear ratios increases, the vehicle is getting closer to a constant power delivered. Describing the maximum acceleration as a function of some physical quantities for relatively large speed, we see that
\begin{equation*}
a_{\text{max}}(v,0) = \frac{P_{\text{wheel}}}{m_{\text{vehicle}}} \left(\frac{1}{v}-\frac{R_a}{P_{\text{wheel}}} v^2\right)\;,
\end{equation*}
where $P_{\text{wheel}}$ is the power at the wheel, $m_{\text{vehicle}}$ is the mass of vehicle, and $R_a$ is the air drag coefficient. If the speed is small, resistance from the air drag is not important, so a lower bound is applied to $a_{\text{max}}(v,0)$ denoted by the parameter $b_1$. Based on visual inspection in the nonzero road grade regime, we observe that $a_{\text{max}}(v,\theta)$ is just a shift in $\theta$ that is piecewise linear in $v$.
We conduct the fitting in two steps:
\begin{itemize}
    \item \underline{Step 1:} Find the parameters $b_1$, $b_2$, $b_3$ by fitting $a_{\text{max}}(v,0)$ to maximum feasible acceleration data collected from the semi-principled model on $500$ equispaced grid points for $v\in[0\text{m}/\text{s}, 70\text{m}/\text{s}]$. 
    
    \item \underline{Step 2:} Find the parameters $b_4$, $b_5$, $b_6$ by first collecting maximum feasible acceleration data from the semi-principled model on 200$\times$150 equispaced grid points for $v\in[0\text{m}/\text{s}, 35\text{m}/\text{s}]$ and $\theta\in[-0.03\text{rad}, 0.03\text{rad}]$. For each fixed speed, we calculate the difference of maximum accelerations for two opposite road grades and use this difference to fit $2\min(b_4, b_5+b_6v)\cdot \theta$ by the least squares method in ($v$,$\theta$)-space. The reason why we multiply by $2\theta$ to the fitted formula rather than using finite differences is to avoid the impact of small $\theta$ near zero.
\end{itemize}
The corresponding fitted parameter results for all vehicle classes can be found in Table~\ref{table:summary-of-fitted-parameters}. A graph illustrating the contours of fuel rate in the feasible region of ($v$,$a$)-space at $0.01\text{rad}$ road grade for both semi-principled and simplified models for a Compact Sedan can be seen in Figure~\ref{fig_feasible_upperbound_SPM_SM}.

\subsection{Physics-like Properties of the Simplified Model}
\label{subsec:physic-like-properties}
One of the desirable properties that were listed for the simplified model in previous section was ``convexity". Here we define precisely what is meant by that. The energy function described in \Sref{subsec:covex_model} is \emph{not} convex in the sense of having a positive semi-definite Hessian matrix. The contours in Figure~\ref{fig_feasible_upperbound_SPM_SM} show also that the function is not convex in ($v$,$a$)-space. Instead, in this study we refer to ``convexity" as the concepts that matter for obtaining a constant speed solution as the fuel-optimal trajectory under reasonable constraints. The reason for highlighting and enforcing that notion of ``convexity" is that the semi-principled model energy function with discontinuities at gear shifting fails to yield that the fuel-optimal trajectory is constant-speed driving---while the simplified model does satisfy it.
In that sense, ``convexity" implies properties of physics-based energy models. The simplified model $f_{\text{s}}(v,a,\theta)$ is actually a generalization of basic physics models \cite{Gong2018} similar to 
\begin{equation*}
f_{\text{b}}(v,a,\theta)= \max(c_0 + c_1 v + c_2 v^2 + c_3 v^3+m_{\text{vehicle}}va+m_{\text{vehicle}}g\theta v,0)\;.
\end{equation*} 
Clearly, the bi-variate term $m_{\text{vehicle}}va$ is a non-convex function. However, considering the simple case of flat roads, this term integrates away when finding the energy-minimizing speed profile for a trajectory on the time interval $[0,T]$.
Thus, ``convexity" here refers to satisfying the following conditions: $\partial_{aa} f_{\text{s}}(v,a,\theta)\geq 0$ and $\partial_{vv} f_{\text{s}}(v,a,\theta)\geq 0$.

\section{Application of the Model Reduction Pipeline: Six Specific Vehicle Models}
\label{vehicle_portfolio}
Here we showcase the above described model reduction pipeline on a portfolio of six vehicle types. This produces, as a contribution of this work, simplified energy models for six vehicle classes; those models are given by simple explicit formulas and thus can easily be included directly into software and publications. Yet, the models are suitably close (see \Sref{sec:validation}) to the Autonomie ground truth, and thus can leverage the accuracy of Autonomie relative to real vehicles.

\subsection{Vehicle Portfolio}
\label{subsec:vehicle_portfolio}
Two principal factors are prioritized in building the vehicle portfolio: diversity and ease of reproducibility. 
In this study, six vehicle models with conventional powertrains are chosen from light-, medium-, and heavy-duty classes. These models are available out of the box in the Autonomie library. The vehicle models chosen do not precisely represent all vehicles in their class due to differences within a class in characteristics that affect energy consumption such as mass, powertrains, technologies, and age. However, the models are chosen to represent an average vehicle of their class without defeating the purpose of simplifying fleet energy consumption estimation by having too many models. Table~\ref{table:model_list} lists the vehicle models in the portfolio and compares some of their characteristics. 

\begin{table*}[htbp]
\begin{center}
\small
\begin{tabular}{|l|l|l|l|l|l|l|}
\hline
Vehicle model & ignition type & duty & maximum  & \#{}gears & total  & frontal  \\
 &  &  & \ power/$\text{kW}$ &  & mass/$\text{kg}$ & area/$\text{m}^2$ \\
\hline
\hline
Compact Sedan     & Spark  & Light       & 99    & 6 & $1379$    &  2.25 \\ 
\hline
Midsize Sedan     & Spark   & Light      & 108   & 6 & $1743$    &  2.25 \\ 
\hline
Midsize SUV       & Spark   & Light      & 99    & 6 & $1897$    &  2.25 \\ 
\hline
Midsize Pickup    & Spark   & Light      & 99    & 6 & $2173$    &  3.27 \\ 
\hline
Class4 PND        & Compression & Medium  & 225   & 5 & $6842$    &  7.50 \\ 
\hline
Class8 Tractor    & Compression  & Heavy  & 285   & 10 & $25104$   &  10.00 \\ 
\hline
\end{tabular}
\end{center}
\vspace{-1em}
\caption{Key characteristics of the chosen portfolio of six representative vehicle models, taken from from Autonomie (PND = Pickup-and-delivery).}
\label{table:model_list}
\end{table*}

\subsection{Specific Vehicle Models}
\label{subsec:specific_vehicle_models}
The semi-principled model described in \Sref{structure_semi_model} is constructed for all six representative vehicles in our portfolio  (\Sref{subsec:vehicle_portfolio}). The obtained semi-principled models are then pushed through the simplification pipeline, and the corresponding fitted models described in \Sref{subsec:covex_model} are established. Below in Table~\ref{table:summary-of-fitted-parameters} is a summary of the fitted parameters for the six vehicle models.

\begin{table*}
\begin{center}
\small
\begin{tabular}{|l|l|l|l|l|l|l|l|l|l|}
\hline 
 & Units& Compact & Midsize & Midsize   & Midsize  & Class4 & Class8\\
 &  & Sedan  & Sedan  & SUV & Pickup &  PND &  Tractor \\ 
\hline 
\hline 
$v_c$ & $\text{m}/\text{s}$ & 5.040e+00 & 5.070e+00 & 9.160e+00 & 1.100e+01 & - & - \\ 
\hline 
$a_0$ & $\text{m}/\text{s}^2$ & -2.698e-01 & -1.574e-01 & -2.685e-01 & -2.646e-01 & - & - \\ 
\hline 
$a_1$ & $1/\text{s}$ & -2.400e-03 & -3.788e-04 & -1.527e-03 & -1.382e-03 & - & - \\ 
\hline 
$a_2$ & $\text{m}\cdot\text{rad}/\text{s}^2$ & -9.062e+00 & -9.112e+00 & -9.430e+00 & -9.494e+00 & - & - \\ 
\hline 
$a_3$ & $1/\text{m}$ & -2.922e-04 & -2.296e-04 & -3.284e-04 & -3.981e-04 & - & - \\ 
\hline 
$a_4$ & $1/(\text{s}\cdot\text{rad})$ & -1.190e-02 & -1.156e-02 & -5.382e-03 & -4.408e-03 & - & - \\ 
\hline 
$\beta$ & $\text{g}/\text{s}$ & 9.720e-02 & 1.271e-01 & 1.637e-01 & 1.999e-01 & - & - \\ 
\hline 
$h_0$ & $\text{g}/\text{s}$ & - & - & - & - & 0 & 4.911e-01 \\ 
\hline 
$h_1$ & $\text{g}/\text{m}$ & - & - & - & - & 3.172e-02 & 2.515e-02 \\ 
\hline 
$c_0$ & $\text{g}/\text{s}$ & 1.592e-01 & 1.983e-01 & 2.250e-01 & 2.632e-01 & 2.429e-01 & 5.945e-01 \\ 
\hline 
$c_1$ & $\text{g}/\text{m}$ & 1.346e-02 & 2.112e-02 & 2.129e-02 & 2.343e-02 & 3.827e-02 & 8.261e-02 \\ 
\hline 
$c_2$ & $\text{g}\cdot \text{s}/\text{m}^2$ & 0 & 0 & 0 & 0 & 0 & 0 \\ 
\hline 
$c_3$ & $\text{g}\cdot \text{s}^2/\text{m}^3$ & 3.189e-05 & 2.780e-05 & 3.765e-05 & 5.521e-05 & 1.870e-04 & 2.728e-04 \\ 
\hline 
$p_0$ & $\text{g}\cdot \text{s}/\text{m}$ & 4.783e-02 & 2.396e-01 & 1.742e-01 & 2.380e-01 & 6.501e-01 & 2.048e-01 \\ 
\hline 
$p_1$ & $\text{g}\cdot \text{s}^2/\text{m}^2$ & 8.697e-02 & 8.059e-03 & 9.462e-02 & 1.029e-01 & 3.338e-01 & 1.196e+00 \\ 
\hline 
$p_2$ & $\text{g}\cdot \text{s}^3/\text{m}^3$ & 6.825e-08 & 2.774e-03 & 7.135e-04 & 1.259e-03 & 2.552e-03 & 1.912e-02 \\ 
\hline 
$q_0$ & $\text{g}\cdot \text{s}^3/\text{m}^2$ & 2.556e-03 & 0 & 0 & 0 & 3.674e-01 & 0 \\ 
\hline 
$q_1$ & $\text{g}\cdot \text{s}^4/\text{m}^3$ & 1.910e-02 & 5.056e-02 & 2.884e-02 & 3.028e-02 & 4.294e-02 & 1.442e-01 \\ 
\hline 
$z_0$ & $\text{g}/(\text{s}\cdot\text{rad})$ & 1.328e-01 & 2.523e+00 & 2.321e+00 & 3.766e+00 & 2.069e+00 & 8.815e-01 \\ 
\hline 
$z_1$ & $\text{g}/(\text{m}\cdot\text{rad})$ & 7.798e-01 & 7.646e-01 & 7.445e-01 & 6.924e-01 & 3.772e+00 & 1.119e+01 \\ 
\hline 
$z_2$ & $\text{g}\!\cdot\!\text{s}/(\text{m}^2\!\cdot\!\text{rad})$ & 1.973e-03 & 6.021e-03 & 1.307e-02 & 2.378e-02 & 0 & 1.884e-01 \\ 
\hline 
$b_1$ & $\text{m}/\text{s}^2$ & 3.360e+00 & 3.922e+00 & 3.338e+00 & 3.016e+00 & 1.264e+00 & 2.423e+00 \\ 
\hline 
$b_2$ & $\text{m}^2/\text{s}^2$ & 4.160e+01 & 4.899e+01 & 5.346e+01 & 6.038e+01 & 1.496e+01 & 8.446e+00 \\ 
\hline 
$b_3$ & $\text{m}$ & 2.119e-04 & 1.396e-04 & 2.390e-04 & 3.833e-04 & 4.953e-04 & 2.627e-04 \\ 
\hline 
$b_4$ & $\text{m}/(\text{s}^2\cdot\text{rad})$ & 8.936e+00 & 8.904e+00 & 9.185e+00 & 9.133e+00 & 9.569e+00 & 9.739e+00 \\ 
\hline 
$b_5$ & $\text{m}/(\text{s}^2\cdot\text{rad})$ & 3.976e+00 & 6.189e+00 & 8.140e+00 & 8.913e+00 & 7.503e+00 & 8.617e+00 \\ 
\hline 
$b_6$ & $1/(\text{s}\cdot\text{rad})$ & 2.448e-01 & 1.005e-01 & 3.430e-02 & 6.303e-03 & 9.943e-02 & 1.576e-01 \\ 
\hline 
\end{tabular}
\caption{Summary of the fitted parameters of the simplified model for a portfolio of six vehicles. A dash (-) implies that the corresponding field is not relevant to this specific vehicle type (light-duty vs.~medium- and heavy-duty).}
\label{table:summary-of-fitted-parameters}
\end{center}
\end{table*}

\section{Validation of the Model Hierarchy}
\label{sec:validation}
Here we validate the developed semi-principled and simplified models, by quantifying their fuel consumption errors with respect to the Autonomie ground truth on relevant standard flat road EPA and UNECE drive cycles and on constructed constant grade drive cycles.

\subsection{Drive Cycles Used for Validation}
\label{subsec:drive_cycles}
The representative vehicles that were chosen to be modeled were simulated and evaluated under different standard drive cycles that explore the cruising and transient behavior of each of the representative vehicles in terms of fuel consumption. Light-duty vehicle template models in Autonomie were simulated under the following four different standard EPA drive cycles, which generate a baseline data to which the semi-principled and simplified models of a portfolio of vehicles covering broad conditions were validated:
\begin{itemize}
    \item \underline{UDDS}, also denoted FTP-72, models city-type driving and is an abridged version of FTP-75. It consists of a cold start phase and a stabilized phase. FTP-75 also has a hot soak phase, which UDDS does not contain \cite{DriveScheduleEPA}.

    \item \underline{HWFET} (Highway Fuel Economy Test). Unlike UDDS, this highway-typical driving pattern has fewer acceleration and deceleration moments and instead has steady behavior for majority of the time \cite{DriveScheduleEPA}.
    \item \underline{WLTC} (Worldwide Harmonized Light-Duty Test Cycle) consists of four sequences exploring different levels of speed and acceleration patterns. Over a total of $1800\text{s}$, the first $589\text{s}$ are low speed driving. The next $433\text{s}$ are a medium speed phase, followed by a high speed phase in the next $455\text{sec}$ and a high speed phase in the final $323\text{s}$ \cite{WLTC2}.
    \item \underline{US06} is a drive cycle that emphasizes aggressive high speed and acceleration driving behavior, fast speed transitions, and after start up driving behavior \cite{DriveScheduleEPA}.
\end{itemize}

\begin{figure}
\centering
\includegraphics[width=0.99\linewidth]{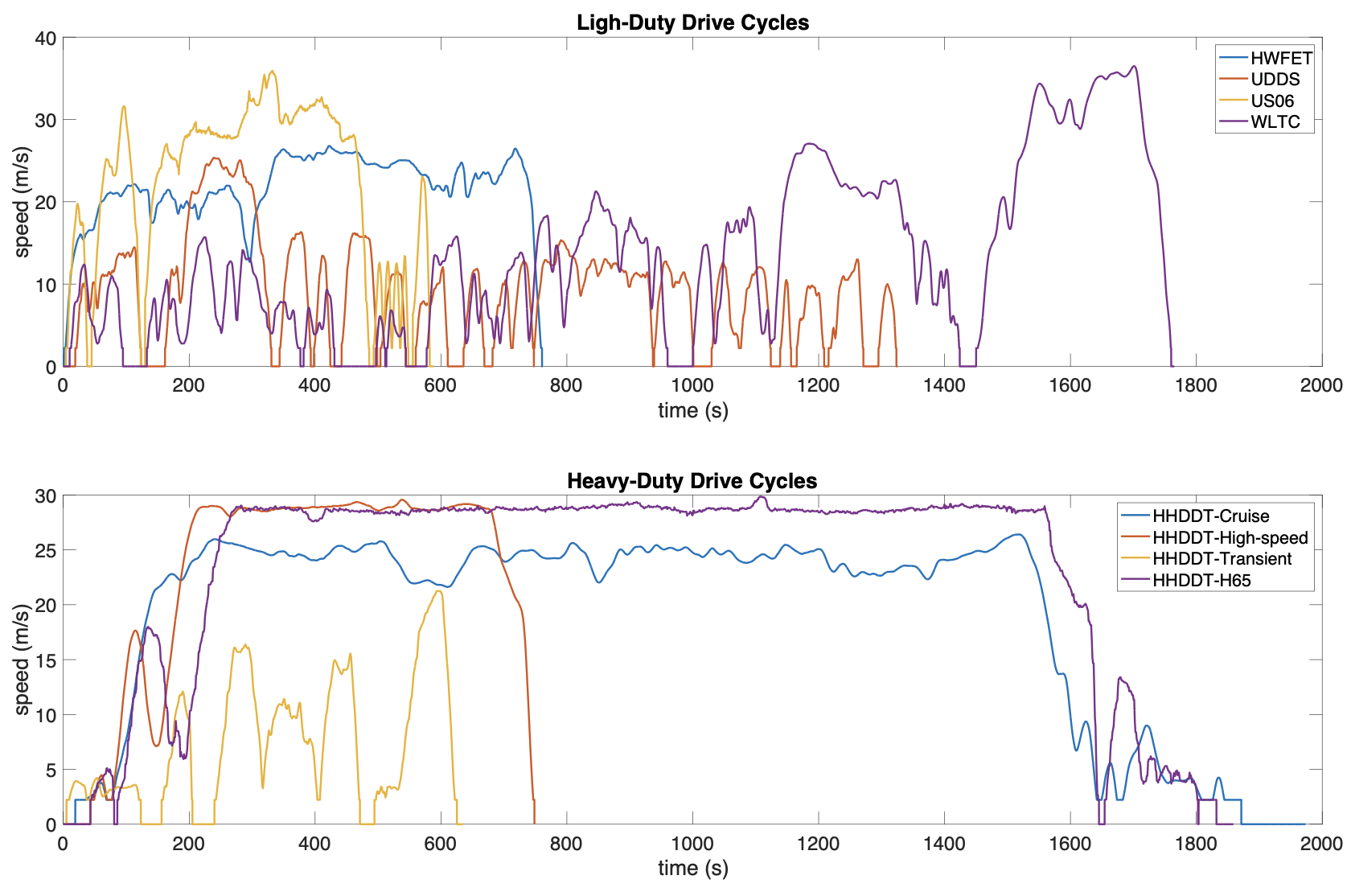}
\caption{Speed profiles for the standard drive cycles used for the validation of reduced energy models for light-duty vehicles (top), and medium- and heavy-duty vehicles (bottom).} 
\label{fig:drive_cycles}
\end{figure}

Shown in Figure~\ref{fig:drive_cycles} are the visualization of these drive cycles. Moreover, after the template vehicle model in Autonomie is modified as needed, it was run under the steady speed cycle to generate the desired Simulink model which will be used for the VCD. 

However, most of these drive cycles are inappropriate for medium- and heavy-duty vehicles as they surpass the speed and acceleration limitation of those vehicles. To resolve this, we replaced the EPA drive cycles with UNECE drive cycles of similar patterns (city and highway driving) that accommodate the limitations of medium- to heavy-duty vehicles. The drive cycles used were the different variants of Heavy Heavy-Duty Diesel Truck (HHDDT), specifically the HHDDT-Cruise, HHDDT-Transient, HHDDT-High, and HHDDT-H65. The HHDDT-Cruise is the equivalent of the city driving pattern for heavy-duty vehicles, while the other three drive cycles are primarily at various steady speeds; the three only differ by the maximum speed and the level and frequency of acceleration and deceleration moments~\cite{HHDDT}.

\subsection{Model Validation on Standard Drive Cycles on Flat Roads}
\label{subsec:validation_flatroads}
\begin{figure}
    \centering
    \includegraphics[width=1\linewidth]{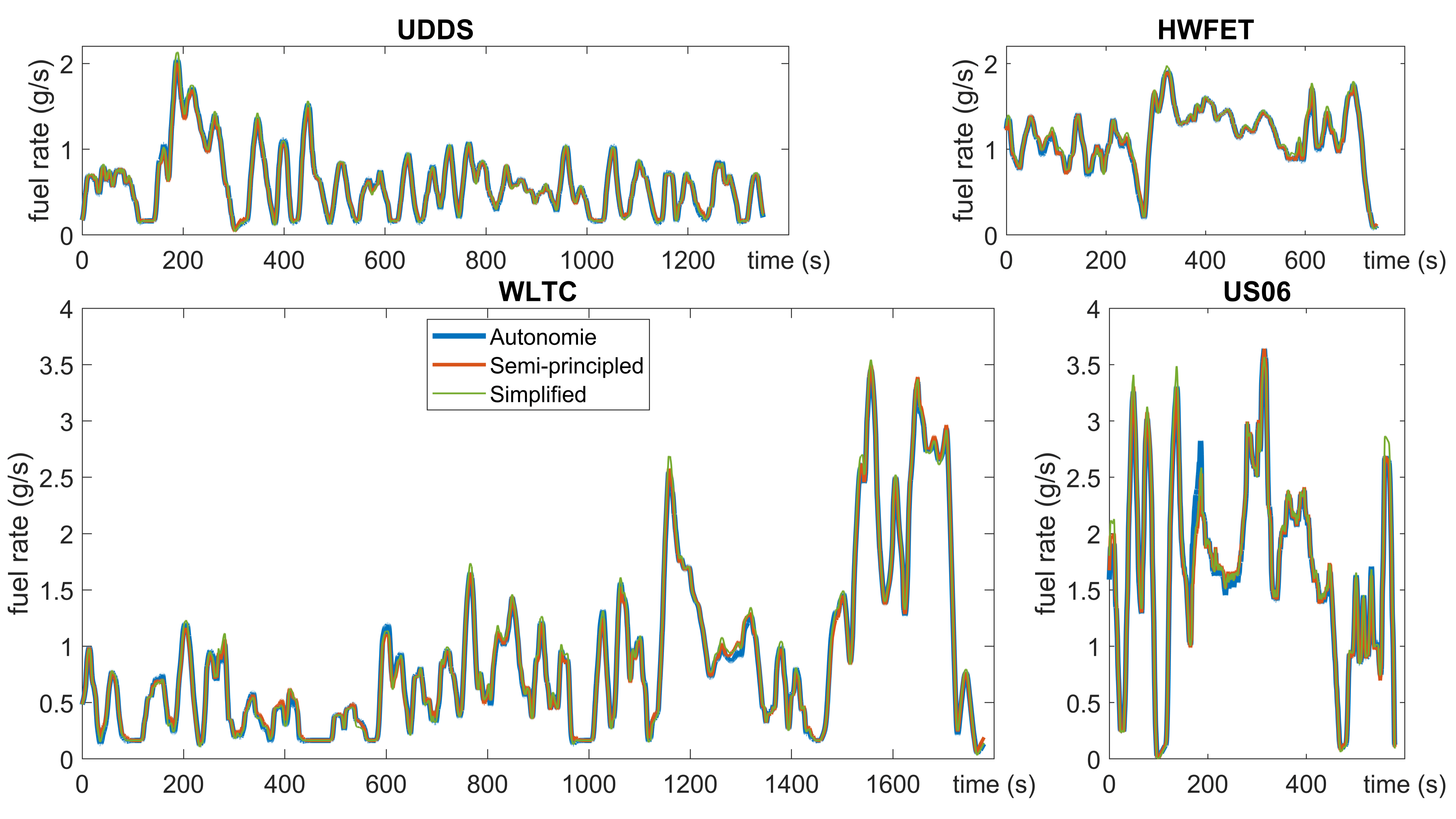}
    \caption{Temporal profiles of the fuel rate in $\text{g}/\text{s}$ for the Midsize SUV, using Autonomie, Semi-principled, and Simplified models on the corresponding light-duty drive cycles. The fuel rate functions are averaged over a $20\text{s}$ moving window across to suppress local-in-time oscillations and capture the effective trend. The plots demonstrated the overall agreement of the reduced models with the ground truth.}
    \label{fig_midSUV_averaged_fuel}
\end{figure}

Here we carry out validations of the semi-principled and simplified models of all six vehicles, relative to Autonomie on the standard EPA and UNECE drive cycles described in \Sref{subsec:drive_cycles}. Table~\ref{table:validation_flat_road} presents the results as percentage relative errors of each model vs.~Autonomie on all the relevant flat road drive cycles combined. For a breakdown of the results drive cycle by drive cycle, see Tables~\ref{table:validation_grade_light} and~\ref{table:validation_grade_heavy}.
To showcase typical plots of the fuel rates over time, Figure~\ref{fig_midSUV_averaged_fuel} shows the fuel rate, averaged over a small moving window, for the Midsize SUV for all three models (Autonomie, semi-principled, and simplified) for the four drive cycles used in the validation (see Appendix \ref{validation_plots} for further validation plots of the complete portfolio of vehicles).

\begin{table*}[htbp]
\begin{center}
\begin{tabular}{|l|l|l|l|l|l|l|}
\hline
  & Compact  & Midsize  & Midsize  & Midsize  & Class4  & Class8  \\
  &  Sedan & Sedan & SUV & Pickup & PND & Tractor \\
\hline
\hline
Semi-principled model & & & & & & \\
relative to Autonomie & -3.7\% & -2.1\% & 0.3\% & 1.2\% & -0.2\% & 0.6\% \\
\hline
Simplified model & & & & & & \\
relative to Autonomie & -1.1\% & 3.3\%& 1.5\% & 0.5\% & 2.0\% & 0.8\%\\
\hline
\end{tabular}
\caption{Summary of the percentage relative errors of the semi-principled models and simplified models vs.~Autonomie on all relevant flat road drive cycles combined for a portfolio of 6 vehicles.}
\label{table:validation_flat_road}
\end{center}
\end{table*}

\subsection{Construction of Drive Cycles with Nonzero Road Grade}
\label{subsec:nonzero_drivecycles}
The standard EPA and UNECE drive cycles described above are all on flat roads. In the absence of standard drive cycles with nonzero road grades, one could construct drive cycles with nonzero road grade starting from those on flat roads. An example is designing an oscillating road grade function $G(t) = 0.03 \sin (\frac{\pi t}{60\text{s}})$ to introduce downhill and uphill shifting throughout the concatenated flat-road drive cycles. However, this would average out the road grade effects when validations are carried. Instead, we construct drive cycles with constant road grades on each EPA and UNECE flat road drive cycle. We choose the maximum and minimum road grades such that the speed, acceleration, and grade combination is feasible for all vehicles on the constructed drive cycles. For light-duty drive cycles the range is $[-0.03\text{rad}, 0.03\text{rad}]$ whereas for heavy-duty drive cycles it is  $[-0.02\text{rad}, 0.02\text{rad}]$.

\subsection{Model Validation on Drive Cycles with Nonzero Road Grade}
\label{subsec:validation-with-grade}
Here we carry out validations of the semi-principled and simplified models, of all six vehicles, relative to Autonomie on the constant nonzero road grade drive cycles constructed in \Sref{subsec:nonzero_drivecycles}. Tables~\ref{table:validation_grade_light} and \ref{table:validation_grade_heavy} present the results as percentage relative errors of each model vs.~Autonomie drive cycle by drive cycle. Errors show that the models do well on flat roads and uphills relative to the ground truth Autonomie. However, errors on downhills are more restrictive because dividing by smaller reference values will generate larger errors. Moreover, downhill driving is harder to capture, specifically in the simplified models, because the fuel-cut dynamics become more relevant, up to (for strong grades) kicking in even at cruising speeds. It should be noted that care must be taken for uphill driving with realizability. Results in light grey denote that the realizabilty is violated, that is, for more than 25\% of the drive cycle time, the Autonomie model fails to reproduce the speed profile with less than $2\text{mi}/\text{h}$ error.

\begin{table*}[htbp]
\begin{center}
\begin{tabular}{|l|l||l|l|l|l|l|l|l|l|l|}
\hline
 \multicolumn{3}{|c|}{Road Grade (rad)} &-0.03 & -0.02 & -0.01 & 0 & 0.01 & 0.02 & 0.03  \\ 
\hline
\hline
\multirow{10}{*}{} &
\multirow{5}{*}{} &
 UDDS & -14.1 & -11.3 & -8.5 & -6.6 & -5.0 & -4.1 & -3.0 \\
\cline{3-10}
 & Semi-principled & HWFET & -8.3 & -5.1 & -3.2 & -1.6 & -1.1 & -0.2 & 1.1 \\
\cline{3-10}
 & relative to & US06 & -2.7 & -2.1 & -1.5 & -1.0 & 0.2 & 0.8 & 1.8 \\
\cline{3-10}
 & Autonomie & WLTC & -6.9 & -5.9 & -4.5 & -3.2 & -2.0 & -1.4 & -0.4\\
\cline{3-10}
Compact &  & ALL & -8.8 & -6.9 & -5.1 & -3.7 & -2.5 & -1.7 & -0.6 \\
 \cline{2-10}
 \cline{2-10}
Sedan & \multirow{5}{*}{} &
 UDDS & -7.4 & -6.1 & -4.9 & -4.1 & -3.6 & -3.3 & -3.0 \\
\cline{3-10}
 & Simplified & HWFET & -5.6 & -1.6 & 0.5 & 1.8 & 1.8 & 1.9 & 2.0 \\
\cline{3-10}
 & relative to & US06 & 1.1 & 0.9 & 0.7 & 0.1 & 0.5 & -0.1 & -0.1 \\
\cline{3-10}
 & Autonomie & WLTC & -2.5 & -1.8 & -1.0 & -0.4 & 0.1 & -0.1 & -0.1 \\
\cline{3-10}
 &  & ALL & -4.1 & -2.7 & -1.7 & -1.1 & -0.7 & -0.7 & -0.6 \\
 \hline
 \hline
\multirow{10}{*}{} &
\multirow{5}{*}{} &
 UDDS & -1.5 & -1.0 & -0.4 & -0.3 & -0.1 & -0.4 & -1.1 \\
\cline{3-10}
 & Semi-principled & HWFET & 0.5 & 0.7 & 0.7 & 0.6 & 0.3 & 0.4 & 0.4 \\
\cline{3-10}
 & relative to & US06 & -0.6 & -0.2 & 0.0 & 0.2 & 0.3 & 0.1 & 0.3 \\
\cline{3-10}
 & Autonomie & WLTC & 1.7 & 0.6 & 0.7 & 0.6 & 0.6 & 0.6 & 0.4 \\
\cline{3-10}
Midsize &  & ALL & 0.2 & 0.0 & 0.3 & 0.3 & 0.3 & 0.2 & -0.1 \\
 \cline{2-10}
 \cline{2-10}
SUV & \multirow{5}{*}{} &
 UDDS & 1.2 & 1.6 & 1.6 & 1.0 & 0.5 & -0.5 & -1.9 \\
\cline{3-10}
 & Simplified & HWFET & -2.3 & 0.7 & 1.9 & 2.2 & 2.0 & 1.6 & 0.8 \\
\cline{3-10}
 & relative to & US06 & 1.7 & 2.1 & 2.1 & 1.7 & 0.7 & 0.2 & -0.4 \\
\cline{3-10}
 & Autonomie & WLTC & 1.1 & 1.3 & 1.7 & 1.6 & 1.4 & 0.7 & -0.5 \\
\cline{3-10}
 &  & ALL & 0.6 & 1.4 & 1.7 & 1.5 & 1.2 & 0.4 & -0.7 \\
 \hline
 \hline
\multirow{10}{*}{} &
\multirow{5}{*}{} &
 UDDS & -6.7 & -5.6 & -4.0 & -3.4 & -3.4 & -3.3 & -4.3 \\
\cline{3-10}
 & Semi-principled & HWFET & 6.3 & 3.5 & 0.8 & -0.2 & -0.9 & -1.5 & -2.0 \\
\cline{3-10}
 & relative to & US06 & -3.1 & -3.0 & -2.5 & -2.4 & -2.6 & -2.3 & -2.0 \\
\cline{3-10}
 & Autonomie & WLTC & -1.7 & -2.1 & -1.6 & -1.8 & -1.8 & -2.1 & -2.7 \\
\cline{3-10}
Midsize &  & ALL & -2.0 & -2.3 & -2.0 & -2.1 & -2.3 & -2.4 & -3.0 \\
 \cline{2-10}
 \cline{2-10}
Sedan & \multirow{5}{*}{} &
 UDDS & 5.8 & 4.5 & 3.8 & 2.6 & 1.4 & 1.1 & 0.6 \\
\cline{3-10}
 & Simplified & HWFET & 17.6 & 11.2 & 6.6 & 4.7 & 3.0 & 1.6 & -0.2 \\
\cline{3-10}
 & relative to & US06 & 8.9 & 5.7 & 4.3 & 2.7 & 0.6 & -0.6 & -2.1 \\
\cline{3-10}
 & Autonomie & WLTC & 7.9 & 6.0 & 5.0 & 3.4 & 2.2 & 1.1 & 0.2 \\
\cline{3-10}
 &  & ALL & 9.0 & 6.4 & 4.8 & 3.3 & 1.9 & 1.0 & -0.1 \\
 \hline
 \hline
\multirow{10}{*}{} &
\multirow{5}{*}{} &
 UDDS & -0.6 & 0.2 & 0.7 & 0.9 & 1.4 & 0.7 & 0.3 \\
\cline{3-10}
 & Semi-principled & HWFET & 0.5 & 1.3 & 1.6 & 1.7 & 1.6 & 1.4 & 1.1 \\
\cline{3-10}
 & relative to & US06 & 0.5 & 0.0 & 0.2 & 0.9 & 1.2 & 1.7 & 1.9 \\
\cline{3-10}
 & Autonomie & WLTC & 1.7 & 1.1 & 1.2 & 1.3 & 1.8 & 0.6 & 0.1 \\
\cline{3-10}
Midsize &  & ALL & 0.6 & 0.7 & 1.0 & 1.2 & 1.6 & 0.9 & 0.6 \\
 \cline{2-10}
 \cline{2-10}
Pickup & \multirow{5}{*}{} &
 UDDS & -2.4 & -1.3 & -0.8 & -0.7 & -0.7 & -1.7 & -2.5 \\
\cline{3-10}
 & Simplified & HWFET & -6.8 & -1.7 & 1.1 & 2.4 & 2.8 & 2.4 & 1.4 \\
\cline{3-10}
 & relative to & US06 & -1.5 & -0.6 & 0.6 & 1.3 & 1.4 & 1.4 & 1.3 \\
\cline{3-10}
 & Autonomie & WLTC & -2.3 & -0.8 & 0.0 & 0.4 & 0.7 & 0.0 & -1.0 \\
\cline{3-10}
 &  & ALL & -3.0 & -1.1 & 0.0 & 0.5 & 0.7 & 0.1 & -0.8 \\
 \hline
\end{tabular}
\caption{Summary of the relative errors (in percent) of the semi-principled and simplified models vs.~Autonomie on flat roads and roads with constant grades for a portfolio of four light-duty vehicles.}
\label{table:validation_grade_light}
\end{center}
\end{table*}

\begin{table*}[htbp]
\begin{center}
\begin{tabular}{|l|l||l|l|l|l|l|l|}
\hline
\multicolumn{3}{|c|}{Road Grade (rad)} &-0.02 & -0.01 & 0 & 0.01 & 0.02 \\
\hline
\hline
\multirow{10}{*}{} &
\multirow{5}{*}{} &
 cruise & 0.2 & -0.2 & -0.1 & 0.6 & 0.8 \\
\cline{3-8}
 & Semi-principled & transient & -2.2 & -1.4 & -0.7 & -1.3 & -1.2  \\
\cline{3-8}
 & relative to & high & 1.2 & 0.5 & -0.1 & 0.1 & \cellcolor{gray}-4.6  \\
\cline{3-8}
 & Autonomie & H65 & 0.6 & -0.1 & -0.2 & -0.4 & \cellcolor{gray}-4.9 \\
\cline{3-8}
Class4 &  & ALL & 0.2 & -0.2 & -0.2 & -0.1 & -2.2 \\
 \cline{2-8}
 \cline{2-8}
PND & \multirow{5}{*}{} &
 cruise & 0.6 & 3.4 & 3.9 & 2.7 & -0.1 \\
\cline{3-8}
 & Simplified & transient & -1.6 & 0.3 & 1.2 & 0.3 & 0.3 \\
\cline{3-8}
 & relative to & high & 2.1 & 2.0 & 0.8 & -0.8 & \cellcolor{gray}-5.9 \\
\cline{3-8}
 & Autonomie & H65 & 0.6 & 1.4 & 0.8 & -1.0 & \cellcolor{gray}-6.2 \\
\cline{3-8}
 &  & ALL & 0.6 & 2.1 & 2.0 & 0.6 & -3.0 \\
 \hline
 \hline
 \multirow{10}{*}{} &
\multirow{5}{*}{} &
 cruise & 0.6 & 0.1 & 1.2 & 0.1 & \cellcolor{gray}1.7 \\
\cline{3-8}
 & Semi-principled & transient & -0.2 & 0.6 & 0.4 & 1.3 & \cellcolor{gray}1.3 \\
\cline{3-8}
 & relative to & high & 3.6 & 1.1 & 0.7 & 2.8 & 3.0 \\
\cline{3-8}
 & Autonomie & H65 & 1.7 & -0.3 & -0.1 & \cellcolor{gray}2.5 & \cellcolor{gray}1.5 \\
\cline{3-8}
Class8 &  & ALL & 1.3 & 0.2 & 0.6 & 1.5 & 1.8 \\
 \cline{2-8}
 \cline{2-8}
Tractor & \multirow{5}{*}{} &
 cruise & 10.8 & -4.1 & 2.1 & 1.1 & \cellcolor{gray}1.5 \\
\cline{3-8}
 & Simplified & transient & -1.7 & -1.0 & -1.8 & -2.2 & \cellcolor{gray}-3.6 \\
\cline{3-8}
 & relative to & high & -3.1 & -3.7 & 0.5 & -0.1 & 0.9 \\
\cline{3-8}
 & Autonomie & H65 & -8.4 & -6.0 & 0.4 & \cellcolor{gray} 0.3 & \cellcolor{gray}1.5 \\
\cline{3-8}
 &  & ALL & 0.4 & -4.3 & 0.8 & 0.3 & 0.8 \\
 \hline
\end{tabular}
\caption{Summary of the relative errors (in percent) of the semi-principled models and simplified models vs.~Autonomie on flat roads and roads with constant grades for a portfolio of two medium- and heavy-duty vehicles. Cells in gray indicate a violation of the realizability criteria described in \Sref{subsec:validation-with-grade}.}
\label{table:validation_grade_heavy}
\end{center}
\end{table*}

\subsection{Discussion: Model Simplicity vs.~Accuracy}
Both the semi-principled and simplified models are designed to be substantially simpler than the complex Autonomie models. Yet, the results above reveal a very good accuracy, with flat-road validation results within a 4\% error margin, averaged over all vehicle classes. That being said, when inspecting the validation results more specifically---drive cycle by drive cycle as well as on non-flat roads---one can see that the models simplicity does lead to compromises in accuracy in certain situations. Here we discuss some of the reasons and modeling choices that ensured structural simplicity but led to increased relative errors, primarily for negative road grades. 

Starting with the semi-principled model, the fitted maps of torque to shaft speed and wheel force assumes a linear function for gears with locked clutch but can underestimate the torque for low speeds where the Autonomie model deviates from the linear relation. This error can become dominant in lower grades with prolonged cruising, such as HWFET, and is mainly recognized for the Midsize Sedan. Moreover, Autonomie takes into account hysteresis effects while the semi-principled and simplified models were designed to be instantaneous models, so this increases errors coming from maps depending on hysteresis such as modeling gear scheduling. 

On the other hand, the simplified models are fittings of the semi-principled models, so errors resulting from the semi-principled model vs.~Autonomie might carry through to errors corresponding to simplified models vs.~Autonomie. Focusing on negative road grades, the amount of fuel consumed there is negligible, so calculating relative percentage errors on negative road grades includes dividing by smaller numbers, and this results in generally larger percentage errors. In the simplified model, the fitting with respect to road grade is done over the domain of positive road grades of interest. This is mainly because we are more interested in uphill driving where more fuel is consumed as compared to downhill. This fitting process ensures a good fit in the downhill regimes as well due to symmetry; however, this symmetry is not sustained for extreme downhills, leading to larger errors there. For all vehicle classes considered, Autonomie's fuel consumption map changes it's general shape for very steep downhills ($<-0.05\text{rad}$). However, this change happens around $-0.02\text{rad}$ for the Midsize Sedan and Class8 Tractor. This explains the larger errors in the simplified model validations for $-0.03\text{rad}$ and $-0.02\text{rad}$ road grades especially for drive cycles with high speed regimes.

\subsection{Semi-principled and Simplified Model Public Release}
The energy models developed herein are publicly released on GitHub and can be found under \cite{circles-energy-models}. The simplified models are provided as fully open MATLAB files for each vehicle. The inputs are instantaneous speed, acceleration, and road grade given as vectors or matrices, resulting in outputs of the same size. Another input is a projection flag (boolean) option, to evaluate the simplified model at the largest feasible acceleration value for the given speed and road grade inputs in case the combination of inputs are not feasible. The outputs are the instantaneous fuel rate and power, along with a feasibility flag that is set to 0 if the inputs are feasible, 1 if the inputs are infeasible, and 2 if the input speed is negative. The semi-principled models are provided in black-box form in the form of 3D arrays of evaluations of the model's output, augmented by MATLAB files with the same structure described above that suitably interpolate the array data. 
Furthermore, a procedure is provided that runs the different models on different drive cycles used for validation to ensure that this work is reproducible.

\section{Conclusions and Outlook}
\label{sec:conclusions}
This work provides a fuel consumption model reduction pipeline from high fidelity software (here: Autonomie) to an approximation into a structurally simpler semi-principled models, and then a further approximation of those models into simplified fitted functions that can be written via simple mathematical polynomials and possess desirable physics-like properties. The key advantages of having such simple yet accurate polynomial models is that they: (i)~are instantaneous in time, i.e, the only inputs are the instantaneous speed, acceleration, and road grade, thus unlike Autonomie they do not require any hysteresis, (ii)~can easily be integrated into real-time simulations or other applications that require fast fuel estimates, (iii)~accurately represent different vehicle types within a vehicle class on the real road as they represent the average vehicle in each class, (iv)~possess desirable properties for optimization and control as they average out local non-convexity behavior due to gear switching to avoid trapping optimizers in local minima, and (v)~are highly accurate with respect to the Autonomie ground truth. The reduction pipeline process is carried out for a portfolio of six vehicle classes: 4 light-duty (Compact Sedan, Midsize Sedan, Midsize SUV, and Pickup) and 2 medium-/heavy-duty vehicles (Class4 PND and Class8 Tractor). The validation of our models on EPA and UNECE drive cycles demonstrated the overall very good agreement of those models with Autonomie as the ground truth, considering the structural simplicity of the new models. Looking forward, we aim at applying this framework to a vehicle-specific model and validating the corresponding semi-principled and simplified models on the results of real chassis dynamometer experiment that we carry out. Moreover, although the presented  framework can in principle be applied to any fuel consumption vehicle model in Autonomie (or any other equivalent software), a natural research step is to consider generalizing the six developed vehicle-class specific models to vehicle type specific (within each class) models via suitable rescalings (based on characteristics like vehicle mass or dimensions) rather than undergoing the whole reduction pipeline process for every vehicle specific model within each class.

\section{Acknowledgments}
The authors would like to thank: (a) the Autonomie team (Argonne National Laboratory), particularly Sylvain Pagerit, Aymeric Rousseau, and Ram Vijayagopal for suggestions and guidance in the usage of the Autonomie software; (b) Sean Murphy (Toyota) for helpful suggestions on the model fitting; (c) Mike Huang (formerly of Toyota) for developing and contributing the initial structure of the semi-principled models used herein and (d) all members of the CIRCLES team who provided helpful feedback on the created models. This material is based upon work supported by the U.S.\ Department of Energy’s Office of Energy Efficiency and Renewable Energy (EERE) under the Vehicle Technologies Office award number CID DE--EE0008872. The views expressed herein do not necessarily represent the views of the U.S.\ Department of Energy or the United States Government. Research was supported by King Abdulaziz City for Science and Technology (S.~Almatrudi). Research was sponsored by the DEVCOM Analysis Center and was accomplished under Cooperative Agreement Number W911NF-22-2-0001. The views and conclusions contained in this document are those of the authors and should not be interpreted as representing the official policies, either expressed or implied, of the Army Research Office or the U.S. Government. The U.S. Government is authorized to reproduce and distribute reprints for Government purposes notwithstanding any copyright notation herein.

\bibliographystyle{plain}
\bibliography{circles_refs, csm_v3}

\clearpage
\nomenclature[O]{\(v\)}{Instantaneous speed}
\nomenclature[O]{\(a\)}{Instantaneous acceleration}
\nomenclature[O]{\(\theta\)}{Instantaneous road grade}
\nomenclature[O]{\(k\)}{Gear}
\nomenclature[P]{\(m_{\text{vehicle}}\)}{Vehicle mass}
\nomenclature[P]{\(m_{\text{general}}\)}{Generalized vehicle mass}
\nomenclature[P]{\(r_{\text{tire}}\)}{Tire radius}
\nomenclature[P]{\(R_{\text{a}}\)}{Air resistance}
\nomenclature[P]{\(R_{\text{r}}\)}{Rolling resistance}
\nomenclature[P]{\(R_{\text{g}}\)}{Frictional load}
\nomenclature[P]{\(d_{\text{r}}\)}{Final drive ratio}
\nomenclature[P]{\(g_{\text{r}}\)}{Gear ratios}
\nomenclature[P]{\(N_{\text{max}}\)}{Maximum engine speed}
\nomenclature[P]{\(N_{\text{min}}\)}{Minimum engine speed}
\nomenclature[E]{\(T_{\text{min}}\)}{Minimum engine torque}
\nomenclature[E]{\(f_{\text{idle}}\)}{Idling fuel rate}
\nomenclature[E]{\(v_{\text{c}}\)}{Fuel-cut speed}
\nomenclature[E]{\(F_{\text{wc}}\)}{Fuel-cut wheel force}
\nomenclature[E]{\(F_{\text{wheel}}\)}{Wheel force}
\nomenclature[E]{\(K_{\text{downshift}}(v)\)}{Downshifting map}
\nomenclature[E]{\(T_{\text{correction}}(a)\)}{Open torque converter correction}
\nomenclature[O]{\(\alpha\)}{Pedal angle}
\nomenclature[S]{\(K_{\text{upshift}}(\alpha,v)\)}{Automatic gear upshift map}
\nomenclature[S]{\(V_{\text{upshift}}(\alpha,k)\)}{Manual gear upshift map}
\nomenclature[S]{\(V_{\text{downshift}}(\alpha,k)\)}{Manual gear downshift map}
\nomenclature[S]{\(T_{\text{max}}(N)\)}{Maximum engine torque map}
\nomenclature[S]{\(T_{\text{wmax}}(v)\)}{Maximum wheel torque map}
\nomenclature[S]{\(T_{\text{max}}(v,k)\)}{Maximum wheel torque map by gear}
\nomenclature[F]{\(f_{\text{r}}(N,T)\)}{Fuel rate map} 
\nomenclature[F]{\(N(N_{\text{output}},F_{\text{wheel}},k)\)}{Engine speed map}
\nomenclature[F]{\(T(N_{\text{output}},F_{\text{wheel}},k)\)}{Engine torque map}
\nomenclature[O]{\(g_{\text{Autonomie}}\)}{Autonomie's gear choice}
\nomenclature[O]{\(N_{\text{output}}\)}{Transmission output speed}
\nomenclature[O]{\(N\)}{Engine speed}
\nomenclature[O]{\(T\)}{Engine torque}
\nomenclature[O]{\(P_{\text{wheel}}\)}{Power at the wheel}
\nomenclature[O]{\(P_{\text{engine}}\)}{Engine power}
\nomenclature[O]{\(f\)}{Semi-principled model fuel rate}
\nomenclature[O]{\(f_{\text{s}}\)}{Simplified model fuel rate}

\setlength{\nomitemsep}{-\parsep}
\printnomenclature

\appendix

\section{Further Validation Plots}
\label{validation_plots}
Above, the model validation results on the drive cycle portfolio were showcased for the Midsize SUV (Figure~\ref{fig_midSUV_averaged_fuel}). Here we display the model validation results over time for the remaining 5 vehicle models on flat roads.
\begin{figure*}[htbp]
    \centering
    \includegraphics[width=1\linewidth]{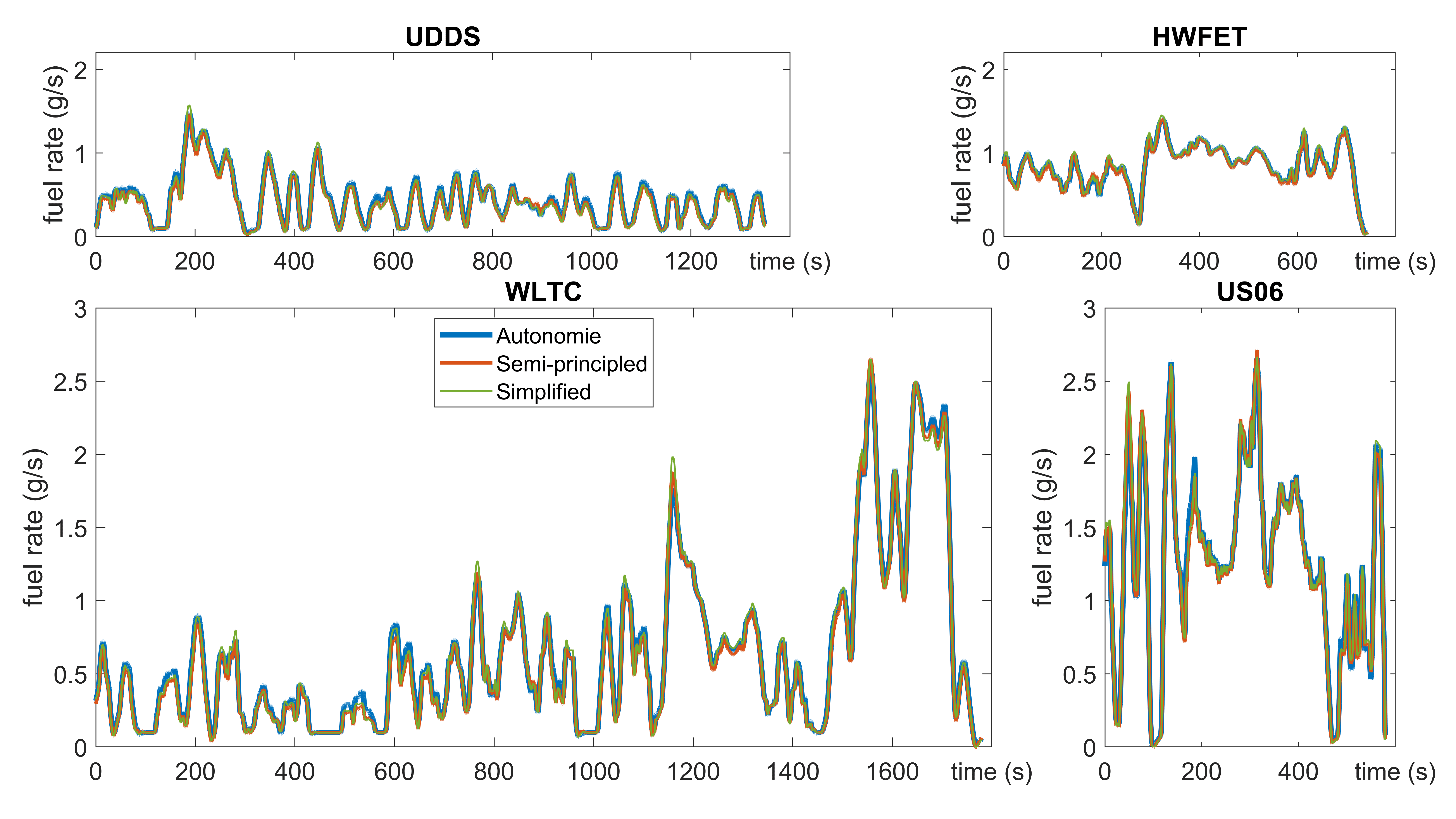}
    \caption{Temporal profiles of the fuel rate in $\text{g}/\text{s}$ for the Compact Sedan, using Autonomie, Semi-principled, and Simplified models on the corresponding light-duty drive cycles.}
    \label{fig_Compact_averaged_fuel}
\end{figure*}

\begin{figure*}[htbp]
    \centering
    \includegraphics[width=1\linewidth]{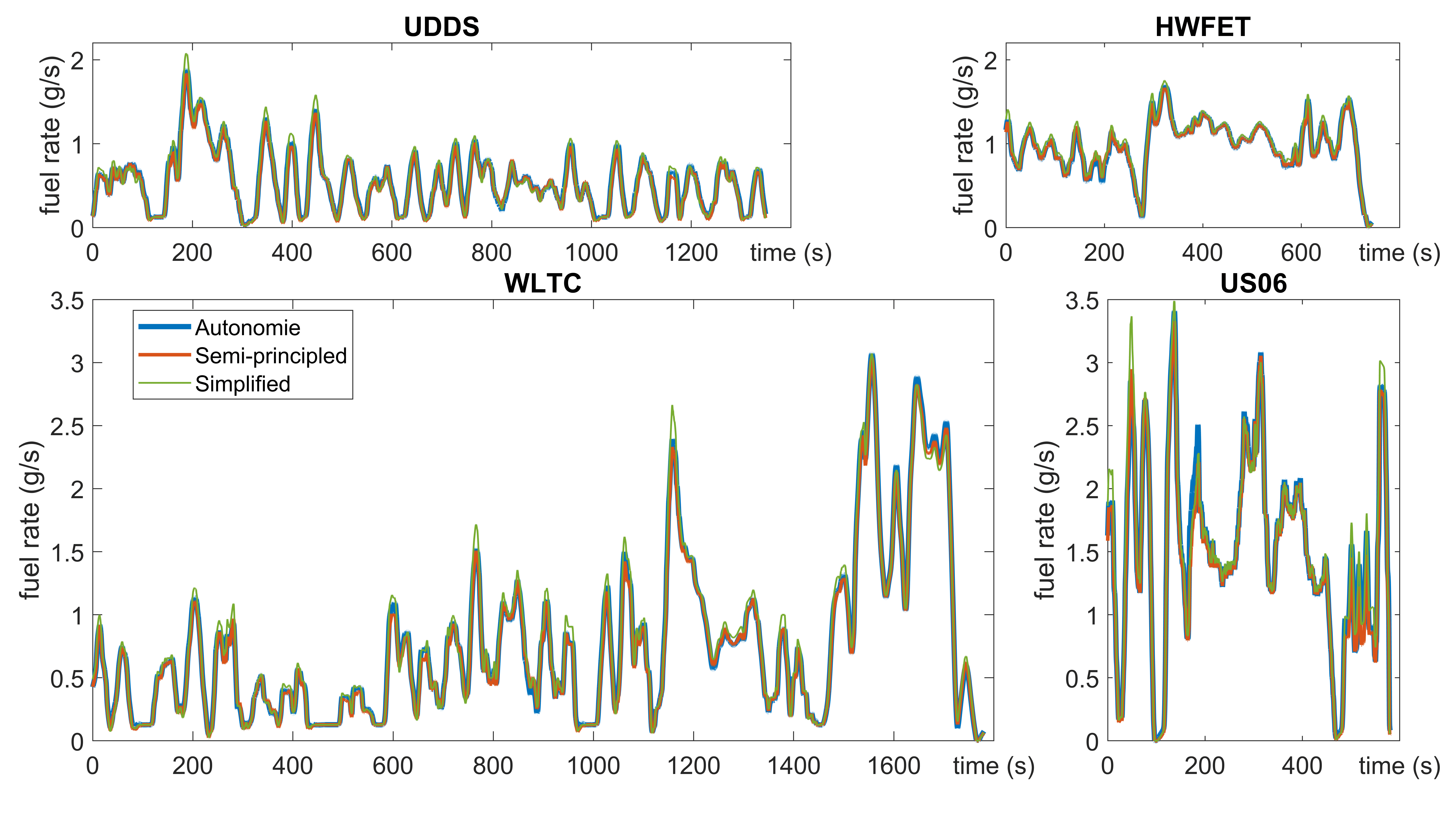}
    \caption{Temporal profiles of the fuel rate in $\text{g}/\text{s}$ for the Midsize Sedan, using Autonomie, Semi-principled, and Simplified models on the corresponding light-duty drive cycles.}
    \label{fig_midBase_averaged_fuel}
\end{figure*}

\begin{figure*}[htbp]
    \centering
    \includegraphics[width=1\linewidth]{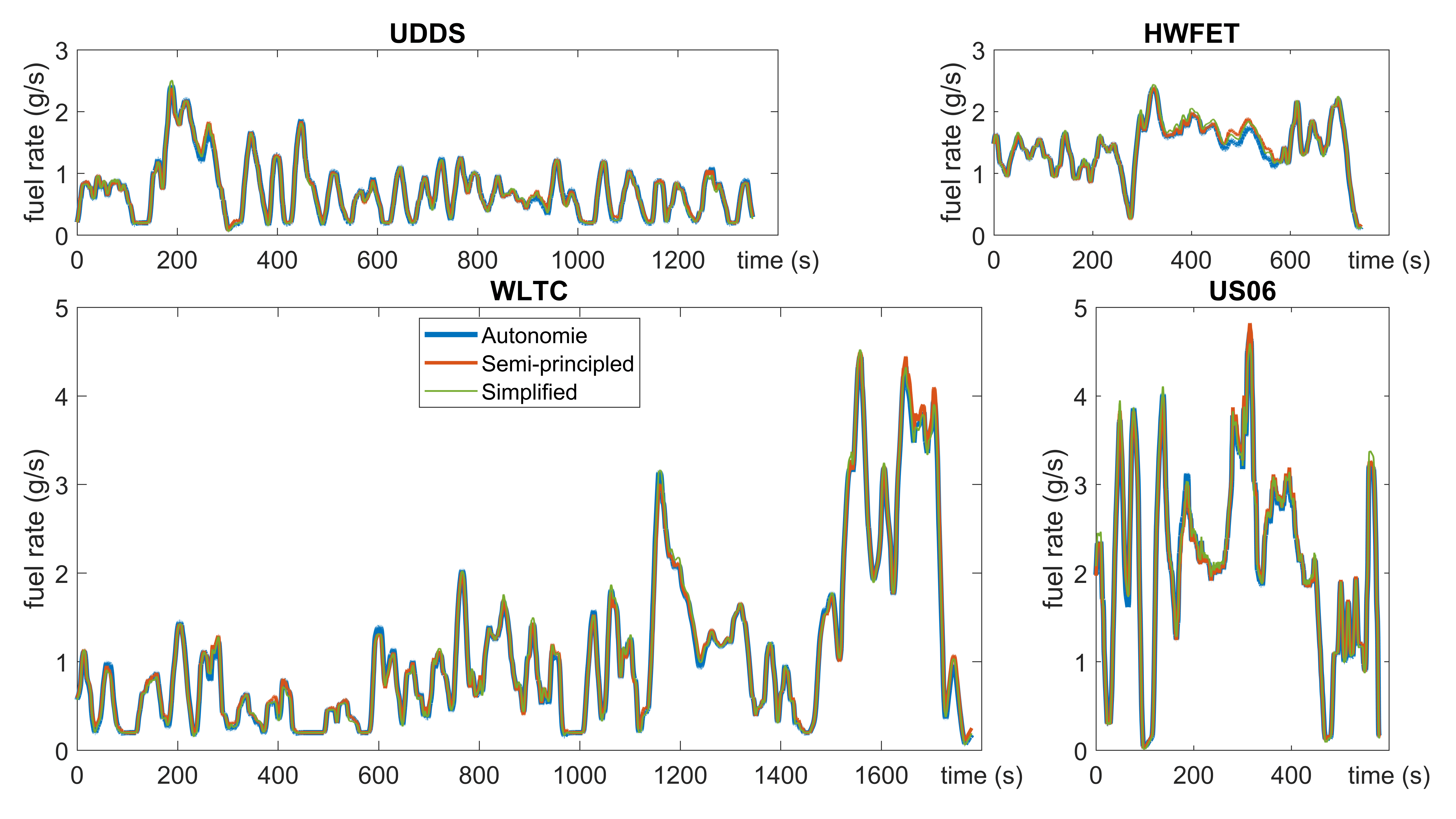}
    \caption{Temporal profiles of the fuel rate in $\text{g}/\text{s}$ for the Midsize Pickup, using Autonomie, Semi-principled, and Simplified models on the corresponding light-duty drive cycles.}
    \label{fig_Pickup_averaged_fuel}
\end{figure*}

\begin{figure*}[htbp]
    \centering
    \includegraphics[width=1\linewidth]{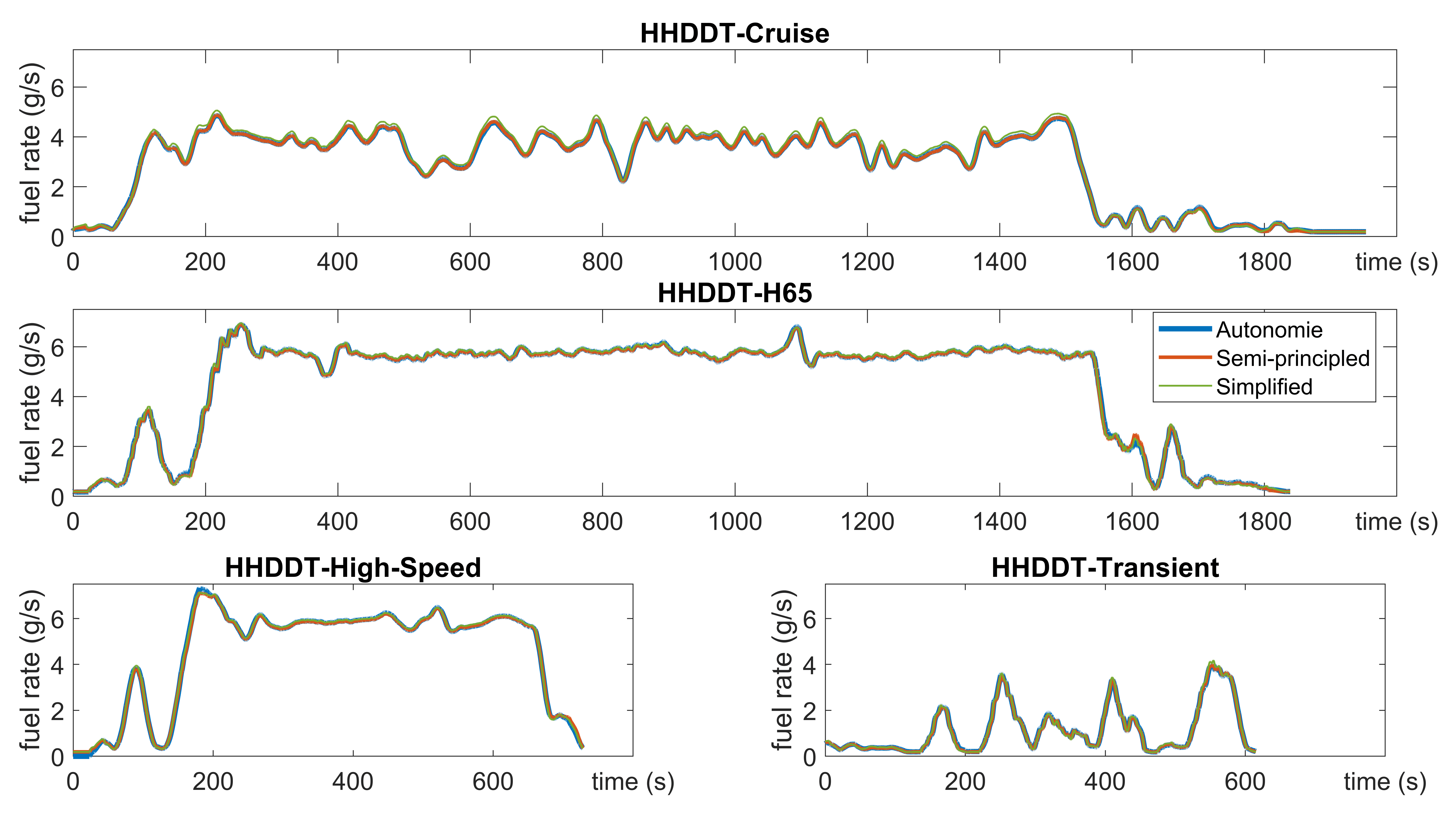}
    \caption{Temporal profiles of the fuel rate in $\text{g}/\text{s}$ for the Class4 PND, using Autonomie, Semi-principled, and Simplified models on the corresponding medium- and heavy-duty drive cycles.}
    \label{fig_Class4PND_averaged_fuel}
\end{figure*}

\begin{figure*}[htbp]
    \centering
    \includegraphics[width=1\linewidth]{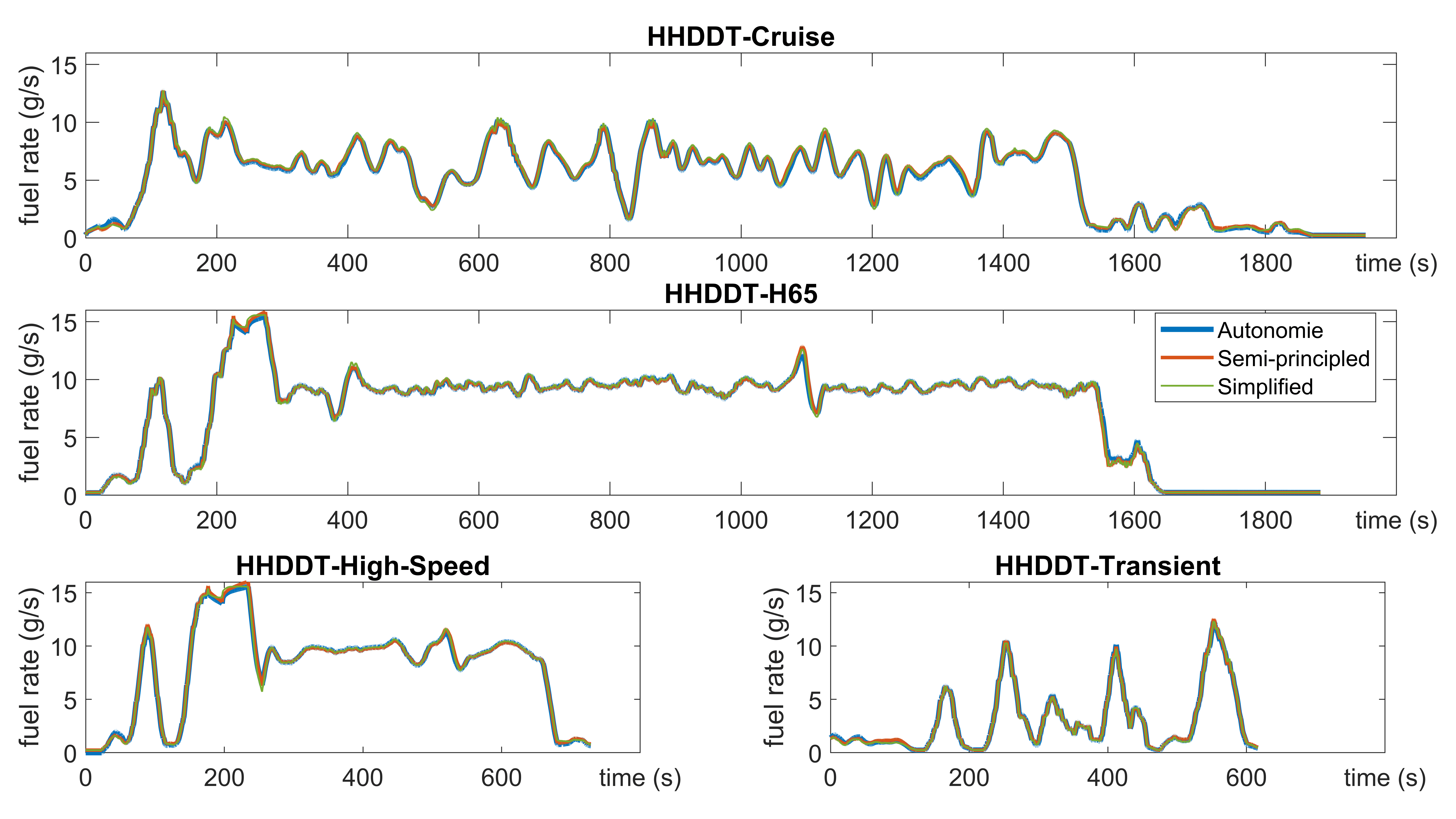}
    \caption{Temporal profiles of the fuel rate in $\text{g}/\text{s}$ for the Class8 Tractor, using Autonomie, Semi-principled, and Simplified models on the corresponding medium- and heavy-duty drive cycles.}
    \label{fig_Class8Tractor_averaged_fuel}
\end{figure*}

\clearpage

\section{Additional Information about the Energy Simulation Tool Autonomie}
Autonomie is a simulation tool that includes Energy Libraries for several vehicle powertrain types that can be used for estimating energy consumption and other vehicle performances such as fuel economy, emissions, regenerative braking, losses from brake, or aerodynamic drag~\cite{ANL}. In this study, Autonomie Rev 16SP7 (MBSE and Vehicle energy library) was used to evaluate energy parameters of  conventional powertrains for the desired vehicle classes. All the baselines used as references and assumptions in the Autonomie compiled vehicles are reported in the Argonne National Laboratory's report to US Department of Energy as part of the 2020 DOE Vehicle Technology Office (VTO) and Hydrogen and Fuel Cell Technology Office (HFTO) baseline and scenario activities \cite{DOE_Report}. Autonomie has more than 100 powertrain configurations for conventional, hybrid (series, parallel, and split), and battery electric types with low-level and high-level controls available for most of the powertrains. There are also several mathematical models for the hardware components of the system (called as the "plant models") with more than 100 initialization modes that can be used to customize a specific vehicle model. A plant model is a MATLAB and Simulink-based model that represents the input-output behavior of several vehicle components starting from (i) Driver model, (ii) Environment model, (iii) High-Level controller of Vehicle Powertrain, and (iv) Vehicle Propulsion Architecture. Models are described in the form of block diagrams with a detailed description of the interconnections between different components. Thus, the model users can explore how the components work together and see which specific input(s) they rely on and what output(s) they can generate. In the Vehicle Propulsion Architecture section can be found the plant and controller models for the engine, clutch, gearbox, chassis and some mechanical and electrical accessories of the vehicles. Pre-defined light, medium, and heavy-duty class vehicle models that are comparable to vehicles currently on the market are available in the Autonomie Library. Thus, Autonomie vehicle model simulation results are considered representative of a vehicle's energy consumption when driven on any of a number of drive cycles. As initial part of the energy modeling, the Autonomie Simulink files were customized in such a way that a virtual chassis dynamometer can be operated into it. Just like how a typical chassis dynamometer works, the Autonomie models were subjected to different drive cycles to which maps like the transmission output speed, wheel force, engine speed and torque and fuel rate were extracted.

\end{document}